\documentclass[twocolumn]{svjour3}          

\smartqed  
\usepackage{graphicx}

\usepackage{wrapfig}
\usepackage{graphicx}
\usepackage{mflogo,texnames}
\usepackage{enumerate}
\usepackage{float}
\usepackage{hyperref}       
\usepackage{url}            
\usepackage{booktabs}       
\usepackage{amsfonts}       
\usepackage{nicefrac}       
\usepackage{microtype}      
\usepackage{times}
\usepackage{xcolor}
\usepackage{soul}
\usepackage{marvosym}

\usepackage{picins}
\usepackage{lineno,hyperref}
\usepackage{enumerate}
\usepackage{amsfonts}       
\usepackage{graphicx}  
\usepackage{subfigure}
\usepackage{multirow}
\usepackage{amsmath}
\usepackage{float}
\usepackage{color}
\usepackage[ruled]{algorithm2e}
\usepackage{framed}

\begin{document}

\title{Link-Backdoor: Backdoor Attack on Link Prediction via Node Injection}



\author{Haibin Zheng         \and
        Haiyang Xiong      \and
        Haonan Ma       \and
        Guohan Huang \and
        Jinyin Chen\Letter
}


\institute{\Letter~Corresponding Author \at
                \email{chenjinyin@zjut.edu.cn}
            \and 
            \textit{H.~Zheng} is with the College of Computer Science and Technology, 
            and Institute of Cyberspace Security 
            at Zhejiang University of Technology, Hangzhou, 310023, China. \\
            \textit{H. Xiong}, \textit{H. Ma}, and \textit{G. Huang} are with the College of Information Engineering at  
            Zhejiang University of Technology, Hangzhou, 310023, China. \\
            \textit{J. Chen} is with the Institute of Cyberspace Security and 
            the College of Information Engineering at 
            Zhejiang University of Technology, Hangzhou, 310023, China.\\
            }

\date{Received: date / Accepted: date}

\maketitle 
    \begin{abstract}
\maketitle 
Link prediction, 
inferring the undiscovered or potential links of the graph, 
is widely applied in the real-world. 
By facilitating labeled links of the graph as the training data, 
numerous deep learning based link prediction methods have been studied, 
which have dominant prediction accuracy compared with non-deep methods. 
However,
the threats of maliciously crafted training graph will leave a specific backdoor in the deep model, thus when some specific examples are
fed into the model, it will make wrong prediction, defined as backdoor attack. It is an important aspect that has been overlooked in the current literature. In this paper, we prompt the concept of backdoor attack on link prediction, and propose  Link-Backdoor to reveal the training vulnerability of the existing link prediction methods. Specifically, the Link-Backdoor combines the fake nodes with the nodes of the target link to form a trigger. Moreover, it optimizes the trigger by the gradient information from the target model. Consequently, the link prediction model trained on the backdoored dataset will predict the link with trigger to the target state. 
Extensive experiments on five benchmark datasets and five well-performing link prediction models
demonstrate that the Link-Backdoor achieves the
state-of-the-art attack success rate under both white-box (i.e., available of the target model parameter)
and black-box (i.e., unavailable of the target model parameter) scenarios. Additionally, we
testify the attack under defensive circumstance, and the results
indicate that the Link-Backdoor still can construct successful attack on the
well-performing link prediction methods. The code and data are available at
\url{https://github.com/Seaocn/Link-Backdoor}.
\keywords{Backdoor attack \and Link prediction \and Node injection\and Gradient information \and Defense}
    \end{abstract}

\section{Introduction}

Our lives are surrounded by various graph-structured data \cite{Di22A,Xi22Graph,Pe21Dual}, representing the complex relationships between objects such as social networks \cite{18Belief}, \cite{Ah20publishing}, biological networks \cite{1992Prediction} and communication networks \cite{2002Scale}. Link prediction is defined to predict the undiscovered or potential links by known nodes and structural features. It has attracted increasing attention from real-world applications. For example, in biological networks \cite{Xu18Mu}, link prediction can help to discover currently unknown protein–protein interactions; in transaction networks\cite{Qi14Pe}, link prediction can find potential products that may be of interest to he/she through his/her purchase records. Numerous effective link prediction methods have been proposed~\cite{V17A}. Similarity-based methods~\cite{Luca06In}, \cite{2009Predicting} obtain the similarity between nodes through the graph structure. Path-based methods~\cite{katz1953new}, \cite{2010Link} exploit the local structural information of the graph to predict links. Besides, embedding-based methods \cite{perozzi2014deepwalk}, \cite{grover2016node2vec},  \cite{tang2015line}  direct their efforts at adopting random walking to characterize the nodes. With the rapid development and wide applications of deep learning methods \cite{liu2020generative}, \cite{liu2019spatial}, the graph neural networks (GNNs) methods have shown promising performance in link prediction \cite{Th17Se}, \cite{Mi18Mo}, \cite{Zh17No}, \cite{Sh21Link}. They take advantage of the nonlinear and hierarchical nature of neural networks to capture the potential characteristic vectors of nodes, which makes prediction accuracy better than non-GNNs based methods.

\begin{figure}[htb]
	\centering
	\includegraphics[width=\linewidth]{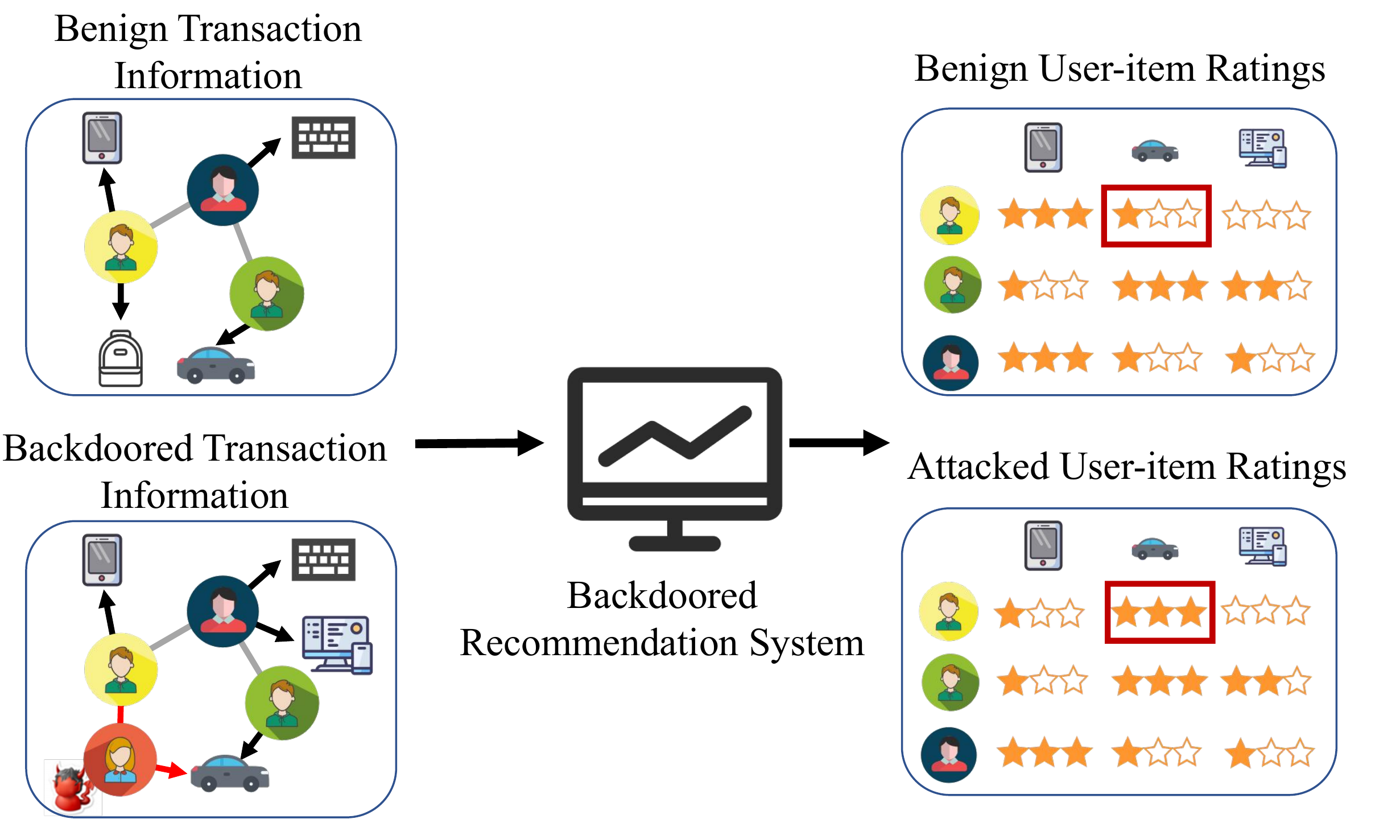}
	\setlength{\abovecaptionskip}{-0.5cm}
	\caption{ An illustration of the backdoor attack on link prediction in the e-commerce scenario.  The red malicious client can mislead the recommendation system by activating the backdoor, leading to the profitable recommendation.}
	\label{fig:real}
\end{figure}

The existing GNNs have achieved satisfactory link prediction performance, benefiting from large number of labeled links in the graph as the training data. However, if the training data is polluted, or even injected with malicious information, it will undermine the link prediction model and bring new security issues. 
Taking the recommendation system as an example, when an e-commerce platform uses benign transaction information to train the recommender system, it can accurately recommend the products that clients will be interested in. However, if a malicious client constructs some fake transaction records and injects them into the transaction information for training, these fake transaction records will leave a backdoor in the recommendation system as Fig. \ref{fig:real}. Then, the malicious client can use the fake transaction records as a trigger to achieve targeted products recommended (e.g., in Fig. \ref{fig:real}, 
the recommendations to yellow user is what malicious client want to recommend  ), 
to gain illegal profits. 
 Consequently, 
 the attacker leaves a backdoor in the model by injecting triggers into the training data, 
 and activates the backdoor to make the link with trigger predicted as the target state during the inference stage, 
 which can be defined as backdoor attack on link prediction model. 
 In this paper, 
 we focus on backdoor attack on link prediction methods to explore the vulnerability of them.

Several backdoor attacks have been proposed against the GNNs on both graph classification  \cite{zhang2021backdoor}, \cite{xi2021graph}, \cite{xu2021explainability} and node classification \cite{zhang2021backdoor}, \cite{xu2021explainability}. Although these backdoor attacks conduct successful attacks on both tasks, they are all challenged in link prediction. In specific, the backdoor attacks on graph classification cannot be applied to link prediction, since it is aimed at the global information of the graph, but it cannot affect the local information of graph such as links in most cases. For backdoor attacks on node classification,  they use specific node features as the
trigger to achieve the backdoor attack. However, the links of graph networks do not have link features, so this kind
of method cannot be used directly.


In summary, the challenges of implementing backdoor attack on link prediction are as follows: (\romannumeral1) \emph{Information Dissemination Limitation}: Inappropriate position of trigger does not effectively affect the target link between the two nodes. (\romannumeral2) \emph{Attacker Limitation}: Modifications of inject node features and trigger structure are limited for stealthiness. (\romannumeral3) \emph{Benign Performance Degradation}: Data manipulation may degrade benign performance in the inference stage. 

To overcome the above challenges, we propose a novel backdoor attack on link prediction, denoted as Link-Backdoor. Specifically, to tackle challenge (\romannumeral1),  we utilize a special subgraph as a trigger, where the two nodes of the target link are part of the trigger (e.g., we can generate triggers by constructing malicious transaction information around the target link), thereby taking the nodes of the target link into account in the trigger, solving the problem that the trigger's features cannot be propagated. To address challenge (\romannumeral2), we use the method of injecting nodes (e.g., the malicious clients) to form triggers. Then, we utilize the gradient information to optimize the structure of the trigger and the features of the injected nodes (e.g., the personal information of the malicious clients). Due to the utilization of injected node and gradient information, the modification of benign data could be minimized. At last, to tackle challenge (\romannumeral3), we use training data with a small amount of triggers to participate in the training of the backdoored model, and reduce the impact on benign performance. 

The contributions of this work are summarized as follows:

\setlength{\hangindent}{2em}
$\vcenter{\hbox{\small$\bullet$}}$ To the best of our knowledge, this is the first work that formulates the backdoor attack on link prediction. 
It reveals the vulnerability of link prediction methods trained on massively collected data, 
and analyzes the possibility of link prediction backdoor attack in practice.

\setlength{\hangindent}{2em}
$\vcenter{\hbox{\small$\bullet$}}$  We design a novel backdoor attack method against link prediction, denoted as Link-Backdoor. 
It uses injection nodes and gradient information to generate and optimize triggers respectively, 
thus it is able to build a relationship between any two nodes in the graph to construct a general attack.

\setlength{\hangindent}{2em}
$\vcenter{\hbox{\small$\bullet$}}$ Extensive experiments on five link prediction methods over five real-world datasets demonstrate that Link-Backdoor can achieve the state-of-the-art (SOTA) attack performance under both
white-box (i.e., available of the target model parameter) and
black-box (i.e., unavailable of the target model parameter)
scenarios. 
Moreover, 
the experiments testify that Link-Backdoor is still effective against possible defense strategy as well.

The rest of the paper is organized as follows. 
Related works are introduced in Section\ref{RWs}. 
The problem definition and threat model are described in Section~\ref{Preliminaries}, 
while the proposed method is detailed in Section~\ref{Methods}. 
Experiment results and discussion are showed in Section~\ref{Exps}. 
Finally, we conclude our work.

\setlength{\parskip}{0\baselineskip}

\section{Related Work\label{RWs}}
In this section, we first briefly review the related work of link prediction
methods. Then, due to the lack of research about backdoor on link prediction, we review the backdoor attacks on graph classification and node classification. Finally, we review the adversarial attack on link prediction and analyze the difference between link prediction adversarial attacks and backdoor attacks.

\subsection{Link Prediction}
Numerous link prediction methods have been proposed. According to the differences in the way of capturing link features, these methods can be divided into four categories: similarity-based methods, path-based methods, embedding-based methods and GNNs-based methods.

\textbf{Similarity-based methods.} The similarity-based methods infer whether two nodes are linked by calculating their similarity. Common neighbors (CN) \cite{newman2001clustering} and resource allocation (RA) \cite{2009Predicting} are two classic similarity-based link prediction methods. They utilize the neighbor information of two nodes to calculate their similarity. Although these methods are simple and effective, they neglect the structural similarity of the nodes in the graph. 

\textbf{Path-based methods.} Consider the paths between two nodes, Katz \cite{katz1953new} proposed an index to calculate their similarity. It sums all the paths between two nodes and assigns more weights to the shorter paths. Liu et al. \cite{2010Link} utilized local random walk (LRW) to capture the transition probabilities of each pair of nodes and used these transition probabilities to measure the similarity between nodes based on LRW.

\textbf{Embedding-based methods.} Due to the good performance of deep learning on natural language processing, Perozzi et al.  \cite{perozzi2014deepwalk} proposed DeepWalk that adopts the random walk and the skip-gram model to train the embedding vector of each node. Based on DeepWalk, Grover and Leskovec \cite{grover2016node2vec} proposed node2vec, which utilizes a biased random walk to generate more flexible node sequences. However, DeepWalk and node2vec only consider the structural information of the graph while ignoring the feature information of the nodes. 

\textbf{GNNs-based methods.} In terms of the nonlinear and hierarchical nature of neural networks, the GNNs-based methods \cite{Th16Va} \cite{Zh17No}, \cite{Mi18Mo}, \cite{Sh18Ad}, \cite{Sh21Link} adopt graph neural networks to capture the potential features vectors of nodes. Based on the local enclosing subgraphs, Zhang and Chen \cite{Mu18Sa} absorbed multiple types of information from subgraph structures and latent node features to learn general graph structure features. Zou and Lerman \cite{Do19En} proposed an encoder–decoder model for graph generation. The encoder is a Gaussianized graph scattering transform, and the decoder can be adapted to link prediction. Motivated by cluster information present in the graph, Mavromatis and Karypis \cite{Co20gr} proposed a graph representation learning method called Graph InfoClust (GIC), that seeks to additionally capture cluster-level information content.

\subsection{Backdoor Attacks on GNNs}
There are currently several studies of backdoor attacks on graph, aiming at graph classification and node classification. Zhang et al. \cite{zhang2021backdoor} proposed a backdoor attack on
graph classification task, which is based on triggers generated by the Erdos-Renyi model. It is designed to establish the relationship between label and trigger of the special structure. Xu et al. \cite{xu2021explainability} further used GNNExplainer to conduct an
explainability research on backdoor attacks of the graph classification and node classification. The other work is graph trojaning attack \cite{xi2021graph}, which is a generative
based method by using a two-layer optimization algorithm
to update the trigger generator and model parameters. Addressing to optimize the random strategy of node selection, Yu et al.\cite{DBLP:conf/colcom/ShengCCK21} proposed a method to evaluate the importance of nodes by combining local and global structural features of nodes and select the most important nodes to form triggers. Yang et al. \cite{DBLP:journals/corr/abs-2207-00425} achieved the first black-box backdoor attack on graph classification by using a surrogate model to train trigger.

The way they choose triggers cannot be directly migrated to the link prediction because of the limitation of trigger propagation. Moreover, they are directly modified the existing link structure or node features of the input data, which is a challenge more difficult to implement in real-world scenarios.

\subsection{Adversarial Attacks on GNNs}
Besides the backdoor attacks on GNNs, there are numerous adversarial attack methods on GNNs. Zügner and Günnemann \cite{DBLP:conf/ijcai/ZugnerAG19} proposed NETTACK that generates the adversarial graph iteratively according to the changing of confidence value after adding perturbations. Based on gradient information, Chen et al. \cite{DBLP:journals/corr/abs-1809-02797} proposed a fast gradient attack (FGA) method that can change the embedding of the target node by modifying a few links. In order to fail the link prediction method, Milani Fard and Wang \cite{DBLP:journals/www/FardW15} proposed a neighborhood randomization mechanism to probabilistically randomize the destination of one link within a local neighborhood.

Although the adversarial attacks and the backdoor attacks both aim to fail the target model, there are still some significant differences.  (\romannumeral1) \emph{Attack Stage}: The adversarial attack launched through the existing vulnerabilities of the target model in the inference stage, while the backdoor attack launched by utilize the learning ability of the target in the training stage. (\romannumeral2) \emph{Model Affection}: The adversarial attack does not affect the model parameters, while the backdoor attack will leave a backdoor in the target model, which will 
modify the model parameters. (\romannumeral3) \emph{Attack Samples}: The backdoor attack can directly utilize preset triggers to achieve attack, while the adversarial attack needs to optimize their samples based on the model's information, such as gradients  \cite{DBLP:journals/corr/abs-1809-02797}, output confidence \cite{DBLP:conf/ijcai/ZugnerAG19}, etc. 

\section{Preliminaries\label{Preliminaries}}
In this section, we introduce the definition of link prediction and backdoor attack on link prediction. For convenience, the definitions of some important symbols used are listed in Table \ref{tab:symbols data}.

\begin{table}[htb]
	\centering
	\caption{THE DEFINITIONS OF SYMBOLS.}
	\label{tab:symbols data}
	\begin{tabular}{r|r}
		\toprule  \hline
		\textbf{Symbols}        &\textbf{Definitions}\\ \hline 
		$G=(V, E)$ & Input graph $G$ with nodes $V$ and edges $E$\\
		$A, X$ & Adjacency matrix, node feature matrix\\
		$\widehat{A}, \widehat{X}$ &  Adjacency matrix / Node feature matrix with triggers\\
		$G$, $\widehat{G}$ & Benign / Backdoored graph \\
		$g$ & Trigger subgraph \\
		$f_{\theta}(\cdot)$ & Benign link prediction model with parameters $\theta$\\
		 $f_{\widehat{\theta}}(\cdot)$ & Backdoored link prediction model with parameters $\widehat{\theta}$\\
		$E_{T}$, $\widehat{T}$ & Target link,  attacker-chosen target link state\\
		$m$, $p$ & Number of injection nodes,  poison ratio\\
		$M(\cdot)$ & Trigger mixture function\\
		$Q_A$, $Q_X$ & Maximum modifications of links and node features\\
		$Gen_{g}(\cdot)$ & Trigger generator\\
		$A_{g}$, $X_{g}$ &   Adjacency / Node feature  matrix of the trigger\\
		$grad_{A_{g}}$ &  Link gradient matrix of trigger\\
		$grad_{X_{g}}$&  Feature gradient matrix of trigger\\
		\hline\bottomrule
	\end{tabular}
\end{table}

\subsection{Problem Definition}
We represent an undirected graph as $G=(V,E)$, where $V$ is the node set, $E$ is the link set. $G$ usually contains an attribute vector of each vertex. Here, we denote the attributes of graph $G$ as $X$ and $A\in \{0,1\}^{n\times n}$ as the adjacency matrix whose element $A_{i, j}=1$ if there is a link between node $i$ and node $j$, otherwise $A_{i, j}=0$. In addition, we use $G = (A, X)$ to represent a graph more concisely. 

\noindent\textbf{Definition 1 (Link Prediction)} For a given graph $G=(V,E)$, $E$ is divided into two groups, $E_{o}$ and $E_{u}$, where $E_{o} \cap E_{u}=\phi$ and $E_{o} \cup E_{u}=E$. $E_{o}$ indicates the set of existing links that are observable. $E_{u}$ indicates the set of links that will be predicted. Link prediction aims to predict missing
links $E_{u}$ based on the information of $V$ and $E_{o}$, where the link prediction method denotes as $f_{\theta}$. 

\noindent\textbf{Definition 2 (Backdoor Attack on Link Prediction)} Given a graph $G$ and the target link $E_T$,  backdoor attack will generate the trigger $g=(A_g,X_g)$ by a  subgraph embedded in training stage. These data with $g$ will participate in model training and leave a backdoor in the backdoor model $f_{\widehat{\theta}}$. In the inference stage, $g$ is  called to make $f_{\widehat{\theta}}$ predict target link $E_T$ as the attacker-chosen state $\widehat{T}$.  Meanwhile, $f_{\widehat{\theta}}$ can still maintain correct predictions on benign data. The adversary’s objective can be formulated as:

\begin{equation}
	\begin{aligned}
		&\left\{\begin{array}{l}
			f_{\widehat{\theta}}\left(\widehat{G}, E_{T}\right)=\widehat{T} , \\
			f_{\widehat{\theta}}(\mathrm{G})=f_{\theta}(\mathrm{G}) ,
		\end{array}\right. \\
		&\textit{s.t.}  \widehat{G}=M(G, g), \left|A_g\right| \leq Q_A,\left|X_g\right| \leq Q_A ,
	\end{aligned}
\end{equation}
where $G$ is the benign graph and $\widehat{G}$ is the backdoor graph with the trigger. $f_{\widehat{\theta}}$ is the backdoored model. $f_{\theta}$ is the benign model. $E_T$ is the target link and $\widehat{T}$ is the attacker-chosen target link state. $M(\cdot)$ is the trigger mixture function. $Q$ is the maximum number of links in trigger. $Q_A$ and $Q_X$ are the maximum modifications of links and node features in trigger.

Intuitively, the backdoor attack on link prediction has two specified goals: (\romannumeral1) it can misclassify the target link of the backdoored graph as the attacker-chosen state; (\romannumeral2) it ensures that benign and backdoored models are consistent in their behavior on the benign graph.

\subsection{Threat Model\label{ThreatModels}}
\noindent\textbf{Attacker’s goal.} An attacker has two goals. First, the attacker should make the link prediction model $f_{\widehat{\theta}}$ predict the target link $E_{T}$ to the attacker's chosen state $\widehat{T}$, when the target link is injected with a trigger $g$. Second, the backdoor attack should not influence the link prediction model's accuracy on benign data, which makes the backdoor attack stealthy.


\noindent\textbf{Attacker’s capability.} We assume two attack scenarios based on the different capabilities of the attacker, including white-box attack and black-box attack. For white-box attack scenario, the attacker not only can obtain part of training data, but also can obtain the structure and parameters of the target model during the training stage. For the black-box attack scenario, the attacker is only allowed to modify part of the training data,  but does not obtain the structure and  parameters of the target model.

\begin{figure*}[htb]
	\centering
	\includegraphics[width=\linewidth]{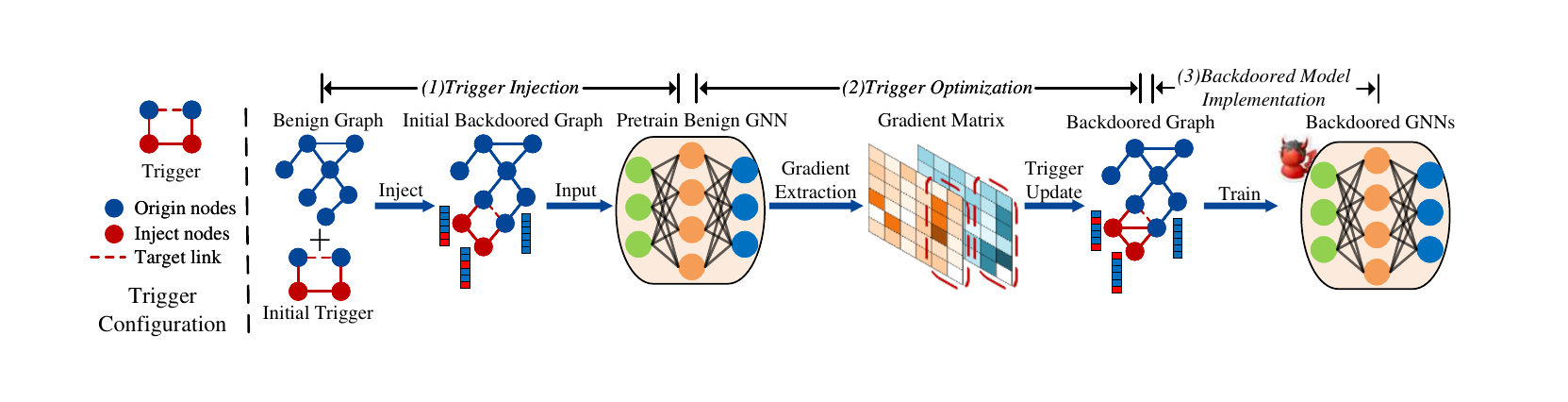}
	\setlength{\abovecaptionskip}{-0.5cm}
	\caption{The overall framework of Link-Backdoor attack. Link-Backdoor launches the backdoor attack on link prediction methods through three steps: 1) Adopting the strategy of injecting nodes and nodes of target link to build the trigger, i.e., a well-designed subgraph; 2) Utilizing the gradient information generate by the link prediction model as a guide to optimize the structure of the trigger and the features of the injected nodes; 3) The training data with triggers will leave a backdoor in the model during the training stage. }
	\label{fig:framework}
\end{figure*}

\section{Methodology\label{Methods}}
As mentioned in Section~\ref{ThreatModels}, 
the Link-Backdoor launches the target link backdoor attack on link prediction model by a carefully designed trigger. In this section, we describe the Link-Backdoor in detail from three steps, i.e., trigger injection, trigger optimization and backdoored model implementation.

The overall framework of Link-Backdoor is shown in Fig. \ref{fig:framework}. First, we inject nodes into the graph and choose any two nodes without link as the target link to form the trigger, and randomly initialize the structure of the trigger. Second, we utilize the gradient information generated by the link prediction model as a guide to optimize the structure of the trigger and the features of the injected nodes. Furthermore, the backdoored data participates in the training of the target model, i.e., the backdoored model.
It is noted that for the black-box scenario, the target model cannot be accessed, so we use a surrogate model to generate the trigger, and inject it into training data for the unknown target model.  
Finally, in the inference stage, the attacker connects the nodes of the target link and the injection nodes according to the topology of the trigger to activate the backdoor, so that the link which is predicted not to exist by the backdoor model is predicted to exist.

\subsection{Trigger Injection}
To realize the backdoor attack that does not modify the existing graph information (e.g., node feature and subgraph topology), we adopt the strategy of the trigger $g$ consisting of injection nodes and target link nodes to avoid modifying existing graph information. Given a benign graph $G=(A,X)$, where $A \in \mathbb R^{N\times N}$ and $X \in \mathbb R^{N\times D}$, we first inject $m$ nodes into the graph $G$ and choose two unlinked nodes as the target link to form the trigger $g$. Second, we randomly initialize the structure of the trigger $g$. 
It can be formulated as:
\begin{equation}\label{equ2}
	g=Gen_{g}(G,m) ,
\end{equation}
where $Gen_{g}(\cdot)$ is the trigger generating function that combines the target link with the injected nodes, $m$ is the number of injection nodes, and these injection nodes can be represented as set $\mathbb{N}_{inj}$. Then, we inject the trigger $g$ into the benign graph $G$, obtaining the backdoored graph $\widehat{G}=(\widehat{A},\widehat{X})$, which can be represented as:
\begin{equation}\label{equ3}
    \begin{array}{c}
    	\widehat{G}=M(G,g) , \\
    	s.t.\left|A_g\right| \leq Q_A   \left|X_g\right| \leq Q_X ,
	\end{array}
\end{equation}
where $M(\cdot)$ is the trigger mixture function that injects $g=(A_g,X_g)$ into a given $G$, where $A_g \in \mathbb R^{(2+m)\times (2+m)}$ and $X_g \in \mathbb R^{(m)\times D}$. $\widehat{G}$ is the backdoored graph, i.e., the graph with the trigger, where $\widehat{A} \in \mathbb R^{(N+m)\times (N+m)}$ and $\widehat{X} \in \mathbb R^{(N+m)\times D}$. $Q_A$ and $Q_X$ are the maximum modifications of links and node features in trigger.

\subsection{Trigger Optimization}
In order to ensure that the backdoor attack is effective, the structure and node features of the trigger need to have an influence on the target link. This means that we need to search for suitable structure and node features for triggers under the perturbation limitation.  An intuitive idea is to search through
permutation and combination, which is extremely
time-consuming. Inspired by the works \cite{DBLP:conf/ijcai/ZugnerAG19}, \cite{DBLP:journals/corr/abs-1809-02797}, they utilize
gradient information to quickly find out nodes or links that are effective for the optimization objective.

Therefore, to generate the effective trigger, we utilize the gradient information generated by the link prediction model to optimize the structure of the trigger and the features of the injected nodes. The trigger optimization is divided into two stages: gradient extraction and trigger update. 

\noindent\textbf{Gradient Extraction.} Specially, given a pre-trained target link prediction model $f_{\theta}$, it already has the ability of link prediction. Then, the backdoored data $\widehat{G}$ is input into the link prediction model $f_{\theta}$ to get the prediction result of the target link. We utilize $L2$ to measure the distance between the predicted state of the target link and the attacker-chosen target link state, which acts as the objective loss function. It can be formulated as:
\begin{equation}\label{equ4}
	\begin{array}{c}
		L_{atk}= \frac{1}{N}\sum_{n=1}^{N}|| f_{\theta}\left(\widehat{G}, E_{T_{n}}\right)-\widehat{T}||_{2}^{2} ,
	\end{array}
\end{equation}
where $E_T$ is the target link. $N$ indicates the number of target links. $\widehat{T}$ is the attacker-chosen target link state and $f_{\theta}$  is the target link prediction model. $|| \cdot ||_{2}$ is the $L2$ distance.

According to this loss function, we can calculate the partial derivative of $L_{atk}$ the structure and features of the trigger $g$. It can be represented as:
\begin{equation}\label{equ5}
	\begin{array}{c}
		grad_{A_{g}}(i,j), grad_{X_{g}}(u,v)=\frac{\partial L_{atk}}{\partial A_{g}(i,j)}, \frac{\partial L_{atk}}{\partial X_{g}((u,v)}  ,\\
	\end{array}
\end{equation}
where $A_{g}$ and $X_{g}$ are the structure and features of the trigger $g$, and  $grad_{A_{g}}$ and $grad_{X_{g}}$ are their gradient matrix in structure and features of the trigger. $i$ and $j$ are the nodes in the trigger and at least one is the injection node, i.e., $(i \cup j) \cap \mathbb{N}_{inj} \neq {\O}$, $u$ is one of the injection nodes, and $v$ is a dimension in the node feature. Considering that the adjacent matrix of an undirected graph is symmetry, we symmetrize $grad_{A_{g}}$ to obtain $grad_{A_{\widehat{g}}}$. 

\begin{equation}\label{equ6}
	grad_{A_{\widehat{g}}}(i,j)= \begin{cases}\frac{grad_{A_{g}}(i,j)+grad_{A_{g}}(j,i)}{2} & ,i \neq j ,\\ 0 & ,i=j , \end{cases}
\end{equation}
where we treat $grad_{A_{\widehat{g}}}$ as a link gradient matrix of the trigger.

\noindent\textbf{Trigger Update.} In the process of backdoor attack, we need to minimize the loss function $L_{atk}$, to let the model prediction close to the selected state. In addition, the positive (or negative) gradient value indicates that the direction of minimizing the target loss is decreasing (or increasing) the value. However, since the graph is discrete, we are only allowed to add or remove links in the trigger. Furthermore, we modify the structure between the trigger according to the structure gradient matrix. Particularly, if the gradient value of any two nodes in the trigger $g$ is positive, we remove the link between the nodes of the trigger $g$, otherwise we add the link between the nodes of the trigger. The optimization of injecting node features is similar to the optimization structure of the trigger. The process can be formulated as:
\begin{equation}\label{equ7}
    \begin{array}{c}
		A_{g}^*(i,j)  =F(A_{g}(i,j) - sign(grad_{A_{\widehat{g}}}(i,j))) , \\
		X_{g}^*(u,v) =  F(X_{g}(u,v) - \alpha(sign(grad_{X_{g}}(u,v)))) ,
		\end{array}
\end{equation}
where $A_{g}^*$ represents the optimized structure of the trigger $g$. $X_{g}^*$ is the optimized features of the injection nodes in the trigger $g$, $\alpha$ is the rate of feature modification, and $\alpha=1$ when the node's feature is 1 or 0. $sign(\cdot)$ is the function that determines whether the effect of the gradient is positive or negative. $F(x)=ReLU(x)-ReLU(x-1)$, which is the function that prevents the final result from exceeding $[0, 1]$.

\subsection{Backdoored Model Implementation}
In this section, we describe in detail how to generate the backdoored model and launch backdoor attack for white-box attack scenario and black-box attack scenario.

For the white-box scenario, we first pre-train the link prediction model $f_{\theta}$ with benign data. It ensures that the link prediction model has correct feedback for the trigger optimizer. Second, we select $N$ target links in the graph and inject the trigger into the target links, i.e., the target links are connected to the injection nodes in the graph. Moreover, we modify the predicted state of the link embedded with the trigger to the attacker-chosen state $\widehat{T}$. Third, we utilize the gradient information generated by the link prediction model to optimize the structure of the trigger and the features of the injected nodes. Finally, the data with the trigger, i.e., backdoored data, participates in the training of the backdoored model $f_{\widehat{\theta}}$. Besides, we take into account that the model parameters will change during the training process, so we will update the trigger every certain training epoch. The details of Link-Backdoor are presented in Algorithm \ref{algorithm_1}.

\begin{algorithm}
\label{algorithm_1}
\caption{Link-Backdoor}
\LinesNumbered
\KwIn{link prediction model $f_{\theta}$,  training Graph $G_{train}$, trigger set $D_{trigger}$, attacker-chosen state of target link $\widehat{T}$, number of model training iterations $O$, trigger update interval $e$, additional nodes $m$}
\KwOut{Backdoored model $f_{\widehat{\theta}}$, trigger $g$} 
Initialization: ${\theta}$  \\
$E_T$ $\leftarrow$ select link from $G_{train}$\\
$g_o$ $\leftarrow Gen_g(G,m)$ by Equation \ref{equ2}. \\ 
Add $g_o$ to $G_{trigger}$.\\
$f_{\widehat{\theta}}$ $\leftarrow f_{\theta}$, $G \in G_{train}$\\
\For {epoch = 1 to $O$} 
{
	\If{epoch $\%$ $e$ == $0$ and epoch $>$ 100}
	{
		$L_{atk}$ $\leftarrow$ calculate the loss of the attack by Equation \ref{equ4}\\
		$grad_{A_g}$ $\leftarrow$ calculate the gradient matrix of structure as Equation \ref{equ5}.\\
		$grad_{A_{\widehat{g}}}$ $\leftarrow$ symmetric the gradient matrix of structure by Equation \ref{equ6}\\
		$grad_{X_g}$ $\leftarrow$ calculate the gradient matrix of feature as Equation \ref{equ5}.\\
		$g$ $\leftarrow$ optimizing triggers by Equation \ref{equ7}.\\
		$\widehat{G}$ $\leftarrow M(G, g)$ by Equation \ref{equ3}.\\
		Add $\widehat{G}$ to $G_{train}$ \\
	}
	$\widehat{\theta}$ $\leftarrow$ update $\theta$ with $G_{train}$ 
}
\Return backdoored model $f_{\widehat{\theta}}$, trigger $g$.
\end{algorithm}

For the black-box scenario, we implement a black-box backdoor attack by transferring backdoored data across the model. First, we adopt one link prediction model as the surrogate model $f_{\theta}$ and train the surrogate model $f_{\theta}$ with benign data. Second, as with the white-box attack, we select $N$ target links and inject triggers into the target links. Moreover, we modify the predicted state of the target link to the attacker's chosen state. Third, we utilize the gradient information generated by the surrogate model to optimize the structure of the trigger and the features of the injected nodes. Finally, we train the target model $f_{\widehat{\theta}}$ with backdoored data optimized by the surrogate model.

After the training is completed, the backdoored model $f_{\widehat{\theta}}$ is obtained. In the inferring stage, the attacker can activate the trigger, i.e., the target link is connected to the injection node in a certain way, to make the backdoored model predict the target link state as the attacker-chosen state $\widehat{T}$.

\subsection{ Theoretical Analysis on Link-Backdoor}
In order to  illustrate the feasibility of our attack method, we conducted a certain theoretical analysis of the Link-Backdoor. We formulate the process by which triggers affect the parameters and output of the model. The theoretical proof verifies that the attacker can leave a backdoor in the model through the trigger. Please refer to the Appendix for details.


\section{Experiments\label{Exps}}
To evaluate our proposed approach, we conduct an empirical study of Link-Backdoor on five benchmark datasets and five SOTA link prediction models. Specifically, our experiments are designed to answer five key research questions (RQs):

\setlength{\hangindent}{2.5em}
\noindent\textbf{RQ1.}
Does the proposed Link-Backdoor achieve the SOTA backdoor attack performance in white-box scenario on link prediction?

\setlength{\hangindent}{2.5em}
\noindent\textbf{RQ2.} Can Link-Backdoor launch successful attack in black-box scenario by transferable poisoning examples?

\setlength{\hangindent}{2.5em}
\noindent\textbf{RQ3.} Can the proposed Link-Backdoor maintain concealment?

\setlength{\hangindent}{2.5em}
\noindent\textbf{RQ4.} Can Link-Backdoor still work well under possible defense?

\setlength{\hangindent}{2.5em}
\noindent\textbf{RQ5.} How do the scale of modified features affect Link-Backdoor? What is the time complexity?

\setlength{\hangindent}{2.5em}
\noindent\textbf{RQ6.} Can Link-Backdoor still attack on non-GNNs methods?


\subsection{Experimental Settings}
\subsubsection{Datasets}
The proposed method is evaluated on five commonly benchmark datasets, i.e., Cora \cite{An00Au}, Cora\_ML \cite{An00Au}, Citeseer \cite{C98Ci}, Pubmed \cite{0Query} and Computer Science (CS) \cite{DBLP:conf/sigir/McAuleyTSH15}. The basic statistics are summarized in Table \ref{tab:data}.

\begin{table}[htb]
	\caption{The basic statistics of five datasets.}
	\centering
	\renewcommand\tabcolsep{5.0pt}
	{
		\begin{tabular}{c|cccc}
			\toprule \hline
			Dataset    &Nodes  &Edges &Classes  & Features  	
			\\ \hline
			Cora \cite{An00Au} &2,708 &5,429 &7 &1,433\\
			Cora\_ML \cite{An00Au} &2,810 &7,981 &7 &2,879\\
			Citeseer \cite{C98Ci} &3,327 &4,732 &6 &3,703 \\
			Pubmed \cite{0Query} &19,717 &44,338 &3 &500\\
			CS \cite{DBLP:conf/sigir/McAuleyTSH15} &18,333 &327,576 &15 &6805\\

			\hline
			\bottomrule
			
		\end{tabular}
	}
	\label{tab:data}
\end{table}

\subsubsection{GNNs-Based Link Prediction Models}
To evaluate the backdoor attack performance
of Link-Backdoor, we choose five SOTA models on link prediction, i.e., GAE \cite{Th16Va}, VGAE \cite{Th16Va}, GIC \cite{Co20gr}, ARGA \cite{Sh18Ad}, ARVGA \cite{Sh18Ad}. We construct a link prediction model with two-layer GCN (i.e., a 32-neuron hidden layer and a 16-neuron hidden layer, while GIC is two 32-neuron hidden layers) and an output layer whose number of neurons is node number. For the rest parameters of the link prediction model, we retain the settings described in the corresponding papers. We briefly describe these models as follows:

\setlength{\hangindent}{2em}
$\vcenter{\hbox{\small$\bullet$}}$ \textbf{GAE \cite{Th16Va}}: This method uses a two-layer GCN as the encoder to obtain the embedding vector, and uses the inner product as the decoder to obtain the reconstructed graph.
It can be formulated as:
\begin{equation}\label{equ101}
	\begin{array}{c}
		Z=\Tilde{A}ReLU(\Tilde{A}XW_0)W_1 ,\\
		\widehat{A}=\sigma(ZZ^T) ,
	\end{array}\\
\end{equation}
where $\sigma$ is the activation function, e.g., Sigmoid, $\widehat{A}$ is the reconstructed graph, $W_0$ and $W_1$ are weight matrices. $\Tilde{A}=D^{-\frac{1}{2}}AD^{-\frac{1}{2}}$  is the symmetrically normalized adjacency matrix, and $D$ is the degree value matrix.

\setlength{\hangindent}{2em}
$\vcenter{\hbox{\small$\bullet$}}$ \textbf{VGAE \cite{Th16Va}}: Compared with GAE, this method adds a variation process when calculating the embedded vector. The calculation process can be formulated as:
\begin{equation}\label{equ102}
	\begin{array}{c}
		Z=\Tilde{A}ReLU(\Tilde{A}XW_0)W_1 + diag({z_i}^2) ,\\
	\end{array}\\
\end{equation}
where $diag(\cdot)$ is the function that preserve the diagonal of the matrix, and $log{z_i}=\Tilde{A}ReLU(\Tilde{A}XW_0)W_1$.

\setlength{\hangindent}{2em}
$\vcenter{\hbox{\small$\bullet$}}$ \textbf{GIC \cite{Co20gr}}: Based on GAE, this method utilizes clustering information and discriminator to improves link prediction performance, it can be formulated as:
\begin{equation}\label{equ103}
	\begin{array}{c}
	    D_K(z_i,c_i)=\sigma({z_i}^Tc_i) ,\\
	\end{array}\\
\end{equation}
where $z_i$ is the node embedding vector output by two layers of GCN, $\sigma$ is activation function, e.g., Sigmoid. $c_i=\sigma (\sum_{k=1}^Kr_{ik}\mu_k)$ is the embedding vector of cluster centers, $r_{ik}$ is the degree that node $n_i$ is assigned to cluster $k$, and $\mu_k = \frac{\sum_i r_{ik}h_i}{\sum_i r_{ik}}$.

\setlength{\hangindent}{2em}
$\vcenter{\hbox{\small$\bullet$}}$ \textbf{ARGA} \cite{Sh18Ad}: This method adds an adversarial training process on the basis of GAE. The adversarial model is built on a standard multi-layer perceptron (MLP), it can be formulated as:
\begin{equation}\label{equ104}
	\begin{array}{c}
	    D=ReLU(ReLU(xw_0)w_1)w_2 ,\\
	\end{array}\\
\end{equation}
where $x$ is the input of the adversarial model, $w_0,w_1$ and $w_2$ are the  weight matrices.

\setlength{\hangindent}{2em}
$\vcenter{\hbox{\small$\bullet$}}$ \textbf{ARVGA} \cite{Sh18Ad}: Similar to ARGA, this method adds an adversarial training process to VGAE. The adversarial model is the same as the adversarial model of ARGA.

\subsubsection{Evaluation Metrics}
To evaluate the effectiveness of the attacks, 
we use two metrics. 
$\left( \romannumeral1\right)$ \emph{attack success rate} (ASR) \cite{Chen20Link}, 
which represents the ratio of the number of incorrectly predicted for the target link to all correctly predicted by the clean model. 
A larger value of ASR indicates better attack performance, 
as follows:
\begin{equation}
	\begin{aligned}
		&\text { ASR }= \frac{\text { Number of successful attack link }}{\text { Number of total attack link }} ,\\
	\end{aligned}
\end{equation}
and $\left( \romannumeral2\right)$ \emph{average misclassification confidence} (AMC) \cite{xi2021graph}, which represents the confidence score of the average output of all successfully attacked links. The higher AMC represents the better performance, which means there is a higher probability of a link existence between the two nodes.

To evaluate the attack evasiveness, we utilize the $\left( \romannumeral1\right)$ \emph{area under curve} (AUC) \cite{Jin05Using}, which represents the accuracy of the link prediction model. If among $n$ independent comparisons, there are $n^{\prime}$ times that the existing link gets a higher score than the nonexistent link and $n^{\prime \prime}$ times they get the same score, then the AUC is defined as:
\begin{equation}
	A U C=\frac{n^{\prime}+0.5 n^{\prime \prime}}{n},
\end{equation}
and based on the AUC, we define the $\left( \romannumeral2\right)$ \emph{benign performance drop} (BPD), which is the AUC difference of two systems built upon the original link prediction model  $f_{\theta}$ and the backdoored link prediction model $f_{\widehat{\theta}}$ with respect to benign inputs. The lower BPD represents the better performance, which suggests that the backdoor attack has a lesser impact on the main performance of the backdoored model.

\subsubsection{Baselines}
Due to the lack of research about backdoor attacks on link prediction, we first transfer two SOTA backdoor attacks in graph classification as baselines, i.e., Erdos-Renyi backdoor (ERB)  \cite{zhang2021backdoor} and graph trojan attack (GTA) \cite{xi2021graph}, to measure the effectiveness of the Link-Backdoor. Moreover, according to the propagation limitation, we combine the nodes of target link with other part remaining nodes to construct the subgraph as the trigger for the two backdoor attacks on link prediction model.

\setlength{\hangindent}{2em}
$\vcenter{\hbox{\small$\bullet$}}$ \textbf{ERB  \cite{zhang2021backdoor}:} ERB generates the trigger by the Erdos-Renyi model, where the probability of each pair of nodes sets 0.8. Then the trigger is injected into the benign dataset for the target model training.

\setlength{\hangindent}{2em}
$\vcenter{\hbox{\small$\bullet$}}$ \textbf{GTA \cite{xi2021graph}:} GTA is a generative backdoor attack. It utilizes a bi-layer optimization  algorithm to update the trigger generator, which generates the trigger with satisfying the constraints. Then the trigger is injected into the benign dataset for the target model training.

Besides, the proposed Link-Backdoor is also compared with its variants (i.e., Random-Backdoor and particle swarm optimization backdoor (PSO-Backdoor)) as baselines. Considering that the topology of the graph is discrete, we choose a discrete optimization method PSO to optimize trigger and implement the backdoor attack.

\setlength{\hangindent}{2em}
$\vcenter{\hbox{\small$\bullet$}}$ \textbf{Random-Backdoor (R-Backdoor):} It randomly selects features from existing node features to act as features for injected nodes. Then, it randomly  generates the link between the injection nodes and the target link nodes. The number of trigger links generated by R-Backdoor is consistent with the number of trigger links of Link-Backdoor.

\setlength{\hangindent}{2em}
$\vcenter{\hbox{\small$\bullet$}}$ \textbf{PSO-Backdoor (P-Backdoor):} It utilizes particle swarm optimization \cite{Chan22Fuzzy} to modify the injection nodes features of the trigger and structure of trigger. We adopt a population size of 200, 50 iterations per round and learning factors of 0.5 and 1.5 in the experiment.

For a fair comparison, triggers generated by different methods participate in the model training process, which is consistent with Link-Backdoor.

\begin{table*}[htb]\large
	\caption{The performance of the five attack methods on the ASR, AMC and BPD of the five link prediction models. We use bold to highlight best attack performance. RB, PB, and LB stand for R-Backdoor, P-backdoor, and Link-Backdoor. }
	\setlength{\tabcolsep}{0.3mm}
	\centering
	\renewcommand\tabcolsep{5pt}
	\resizebox{\linewidth}{!}{
		\begin{tabular}{c|c|c|ccccc|ccccc|ccccc}
			\toprule \hline
			\multirow{3}{*}{Datasets}   &\multirow{3}{*}{Target Models} &\multirow{2}{*}{AUC(\%)}  & \multicolumn{5}{c|}{ASR(\%)} &\multicolumn{5}{c|}{AMC($\times 10^{-2}$)}  & \multicolumn{5}{c}{BPD(\%)}  
			\\ \cline{4-18} & & &\multicolumn{4}{c}{Baselines} &Ours &\multicolumn{4}{c}{Baselines} &Ours &\multicolumn{4}{c}{Baselines} &Ours   
			
			\\\cline{3-18} & &benign &ERB &GTA &RB &PB &LB &ERB &GTA &RB &PB &LB &ERB &GTA &RB &PB &LB \\ \hline
			
			\multirow{5}{*}{Cora} 
			&GAE &90.67 &58.33 &49.55 &77.38 &79.45 &\textbf{81.25} &58.73 &58.45 &59.84 &60.95 &\textbf{61.25} &2.98 &5.87 &4.25 &\textbf{1.42} &2.82  \\ 
			&VGAE &90.10 &61.22 &54.61 &74.11 &79.46 &\textbf{83.22} &58.51 &57.92 &59.36 &61.85 &\textbf{63.47} &10.92 &3.95 &4.77 & \textbf{3.85} & 4.65 \\
			&GIC  &92.18 &33.38 &68.65 &34.52 &96.72 &\textbf{99.41} &55.23 &84.75 &57.57 &65.27 &\textbf{98.24} &1.14 &10.29 &\textbf{0.73} &7.50 &9.92 \\ 
			&ARGA  &92.52 &54.05 &54.46 &66.07 &84.11 &\textbf{92.56} &58.02 &58.19 &56.93 &\textbf{65.71} &62.16 &7.61 &7.04 &8.40 &6.12 &\textbf{1.08} \\  	 	 	 
			&ARVGA &89.77 &71.13 &54.02 &68.75 &69.94 &\textbf{85.25} &59.04 &57.89 &57.01 &58.58 &\textbf{59.29} &5.63 &4.69 &4.21 &\textbf{2.24} &2.49 \\  \hline
			
			\multirow{5}{*}{Cora\_ML}
			&GAE  &91.96 &61.59 &75.91 &72.41 &90.31 &\textbf{90.35} &57.27 &59.25 &57.36 &63.72 &\textbf{64.01} &\textbf{1.38} &5.79 &3.73 &3.82 &3.65\\
			&VGAE &89.63 &82.79 &79.17 &77.03 &85.14 &\textbf{87.54} &61.93 &59.23 &57.79 &63.35 &\textbf{63.69} &5.35 &6.22 &4.41 &3.22 &\textbf{2.91} \\ 
			&GIC &93.51 &42.79 &94.75 &41.98 &99.95 &\textbf{100.00} &55.37 &91.09 &52.67 &99.82 &\textbf{99.95} &\textbf{1.32} &16.77 &1.7 &21.04 &18.10\\ 
			&ARGA  &93.05 &68.84 &69.93 &69.56 &88.76 &\textbf{89.13} &57.84 &59.81 &56.69 &62.96 &\textbf{64.19} &5.42 &8.12 &4.71 &4.86 &\textbf{4.69}\\ 
			&ARVGA  &90.63 &58.88 &71.38 &74.64 &86.82 &\textbf{88.25} &56.43 &59.30 &56.91 &63.32 &\textbf{64.69} &4.32 &7.93 &5.68 &3.44 &\textbf{3.37}\\  \hline
			
			\multirow{5}{*}{Citeseer}
			&GAE &89.65 &60.92 &54.08 &75.60 &81.42 &\textbf{82.85}  &59.70 &56.41 &59.74 &60.47 &\textbf{61.47}  &6.26 &9.05 &4.54 &7.62 &\textbf{3.13} \\
			&VGAE &88.59 &57.55 &76.84 &75.74 &77.92 &\textbf{89.59} &57.53 &59.75 &59.23 &60.73 &\textbf{63.16} &6.90 &8.34 &8.10 &\textbf{3.23} &8.51 \\ 
			&GIC &93.72 &41.94 &50.2 &38.77 &99.64 &\textbf{99.86} &72.54 &78.17 &77.41 &99.73 &\textbf{99.90} &2.01 &11.27 &\textbf{0.30} &10.11 &9.79 \\ 
			&ARGA  &89.67 &46.43 &66.33 &61.10 &93.14 &\textbf{98.78} &58.23 &58.05 &55.32 &65.13 &\textbf{71.33} &5.22 &7.49 &8.53 &\textbf{4.84} &4.97 \\ 
			&ARVGA  &87.63 &55.92 &77.40 &63.67 &\textbf{83.46} &81.55 &57.57 &59.39 &56.11 &59.88 &\textbf{60.01} &4.23 &8.13  &1.87 &2.94 &\textbf{1.60}\\  \hline
			
			\multirow{5}{*}{Pubmed }
			&GAE &95.07 &68.06 &53.24 &66.92 &66.01 &\textbf{76.80} &54.49 &55.53 &54.19 &54.34 &\textbf{55.54} &5.42 &11.25 &6.67 &2.27 &\textbf{1.47} \\  
			&VGAE  &89.05 &65.97 &76.62 &76.43 &79.50 &\textbf{85.21} &54.82 &57.35  &55.34 &\textbf{57.39} &54.50 &0.55 &8.12 &3.36 &0.91 &\textbf{0.20} \\ 
			&GIC  &91.14 &28.81 &50.55 &29.83 &87.22 &\textbf{87.98} &57.72 &57.75 &59.14 &66.11 &\textbf{68.98} &1.88 &7.69 &\textbf{0.55} &6.65 &10.86 \\ 
			&ARGA &94.40 &70.03 &54.04 &69.60 &81.55 &\textbf{83.25} &54.37 &55.45 &54.3 &\textbf{56.43} &55.37 &4.03 &10.55 &4.53 &\textbf{1.47} &2.96 \\ 
			&ARVGA  &90.18 &71.92 &81.96 &75.05 &79.51 &\textbf{84.33} &55.28 &57.40  &56.41 &56.43 &\textbf{57.76} &4.91 &11.35 &3.93 &0.47 &\textbf{0.04 }\\  \hline
			
			\multirow{5}{*}{CS }
			&GAE  &92.98 &33.29 &40.99 &41.22 &41.48 &\textbf{50.61} &55.47 &55.23 &55.52 &54.71 &\textbf{55.85} &3.13 &\textbf{0.23} &3.37 &3.47 &1.13\\  
			&VGAE  &93.66 &56.09 &63.64 &65.37 &65.39 &\textbf{67.26} &55.64 &55.73 &55.58 &55.15 &\textbf{55.94} &2.73 &3.20 &2.50 &5.81 &\textbf{0.95}\\ 
			&GIC   &93.57 &29.91 &27.11 &33.91 &64.67 &\textbf{65.77} &57.27 &54.34 &56.35 &57.57 &\textbf{57.62} &1.30 &7.75 &\textbf{0.01} &1.05 &0.99\\ 
			&ARGA  &91.99 &53.14 &45.73 &47.53 &62.82 &\textbf{63.52} &52.52 &\textbf{55.52} &54.43 &53.57 &54.78 &6.70 &0.35 &7.79 &0.37 &\textbf{0.32}\\ 
			&ARVGA  &93.32 &67.39 &62.47 &57.53 &62.35 &\textbf{68.67} &55.96 &56.15 &56.11 &56.19 &\textbf{56.24} &3.36 &4.54 &8.22 &3.43 &\textbf{2.86}\\  \hline
			
			\hline
			\bottomrule
		\end{tabular}
	}
	\label{tab:main}
\end{table*}

\subsubsection{Experimental Settings}
Following some previous works \cite{Th16Va}, \cite{Co20gr}, we split the data into training, validation, and test sets with a ratio of 85:5:10. We set the number of injected nodes to 2, the maximum number of trigger connections to 5 and the poisoning ratio to 0.1 \textit{(m = 2, Q = 5 and p = 0.1)} in the experiment. We employ the early stopping criterion during the training process, i.e., we stop training if the validation loss does not decrease for 100 consecutive epochs. In the experiment, we repeated the above steps five times and reported the average performance.

\subsubsection{Experimental Environment}
We implement Link-Backdoor with PyTorch, and our experimental environment consists of i7-7700K 3.5GHzx8 (CPU), TITAN XP 12GiB (GPU), 16GBx4 memory (DDR4) and Ubuntu 16.04 (OS). 

\begin{figure}[htbp]
	\centering
	\includegraphics[width=1\linewidth]{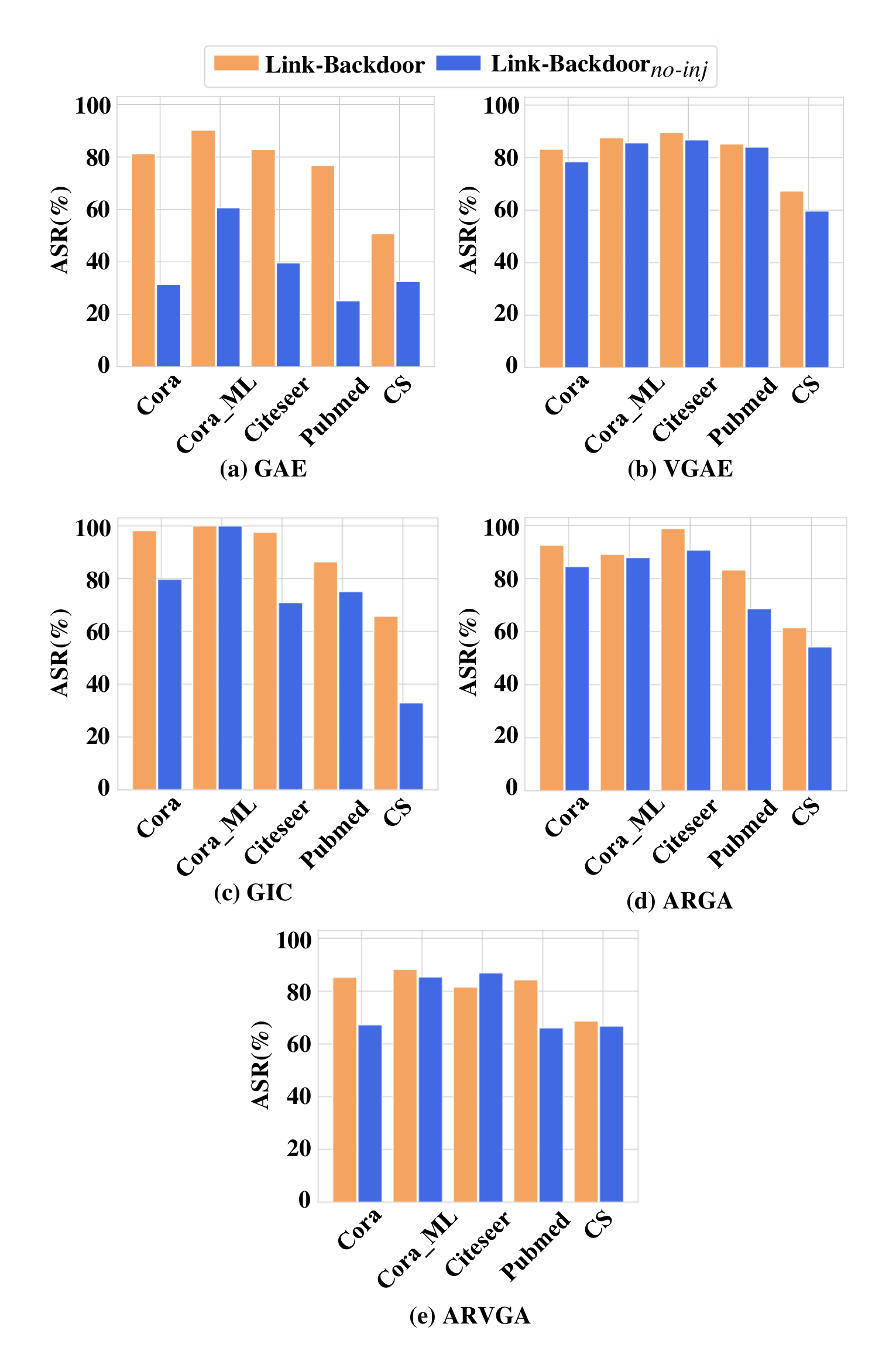}
	\caption{The ASR value for ablation study of Link-Backdoor on five datasets. We choose nodes from the original graph to build the trigger operation to replace the node injection module, expressed as  Link-Backdoor$_{no-inj}$.}
	\label{fig:abl}
\end{figure}

\begin{figure*}[] 
	\centering
	\subfigure[Cora]{
		\includegraphics[width=0.28\linewidth]{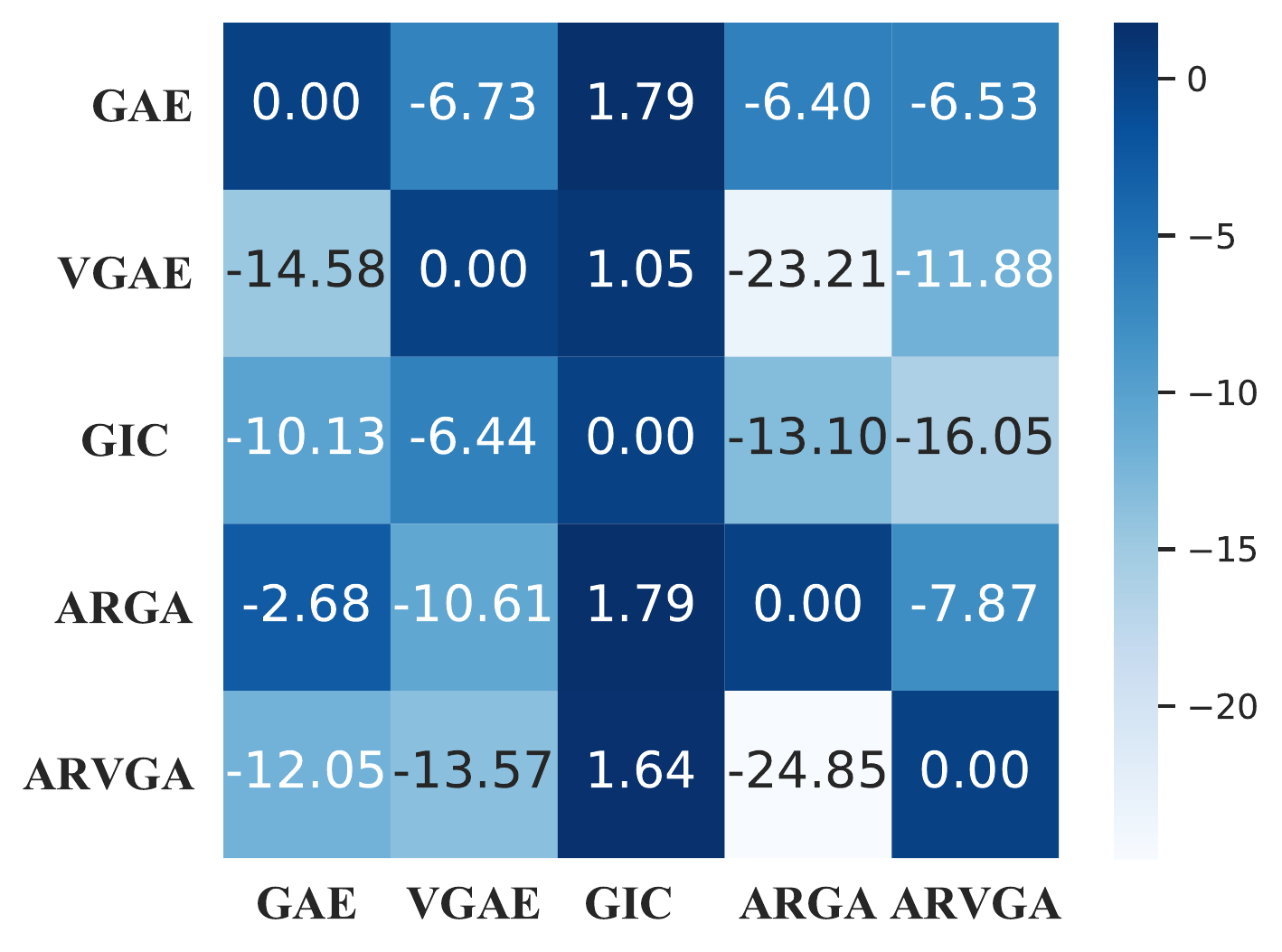}
	}
	\quad
	\subfigure[Cora\_ML]{
		\includegraphics[width=0.28\linewidth]{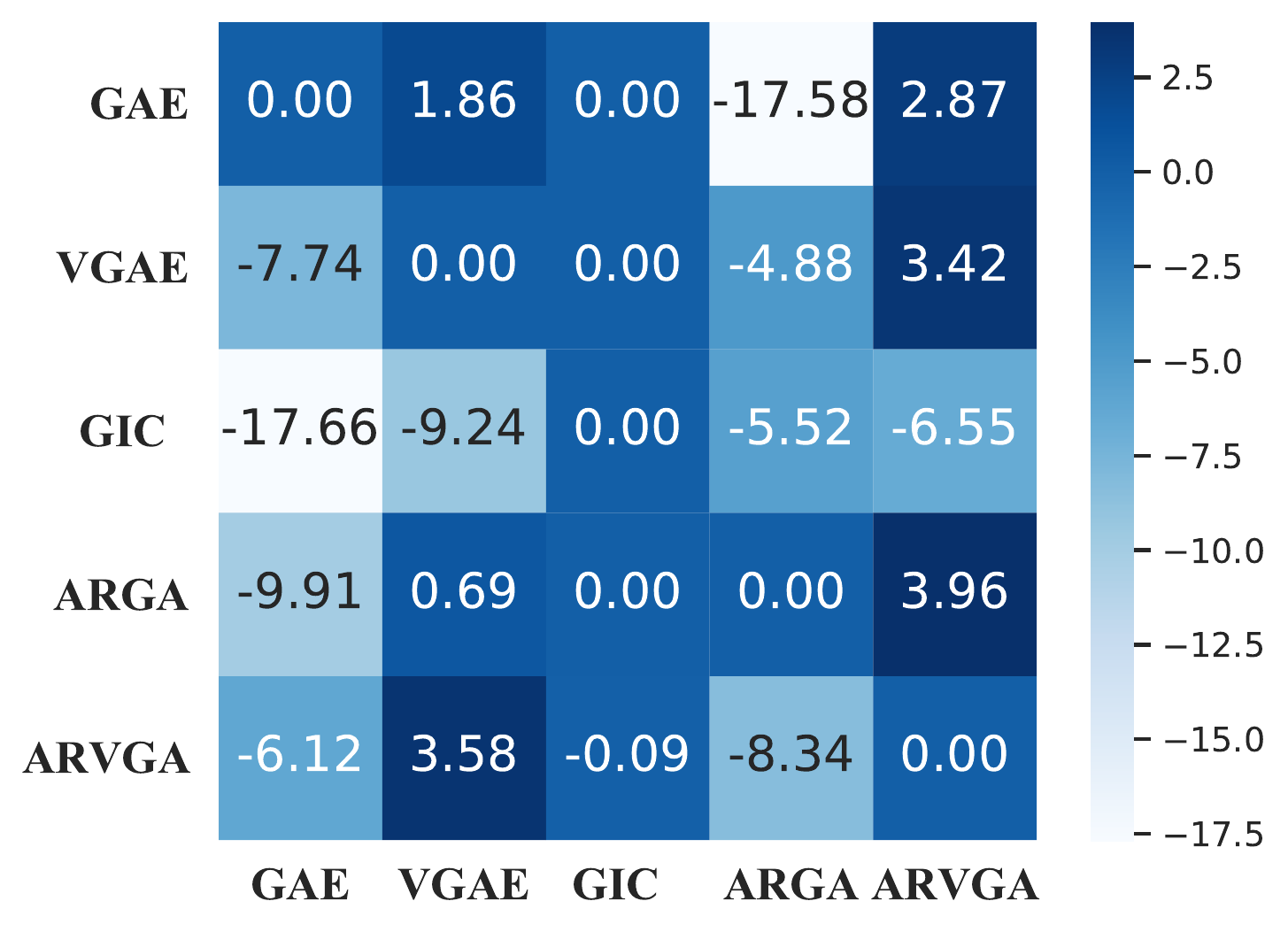}
	}
	\quad
	\subfigure[Citeseer]{
		\includegraphics[width=0.28\linewidth]{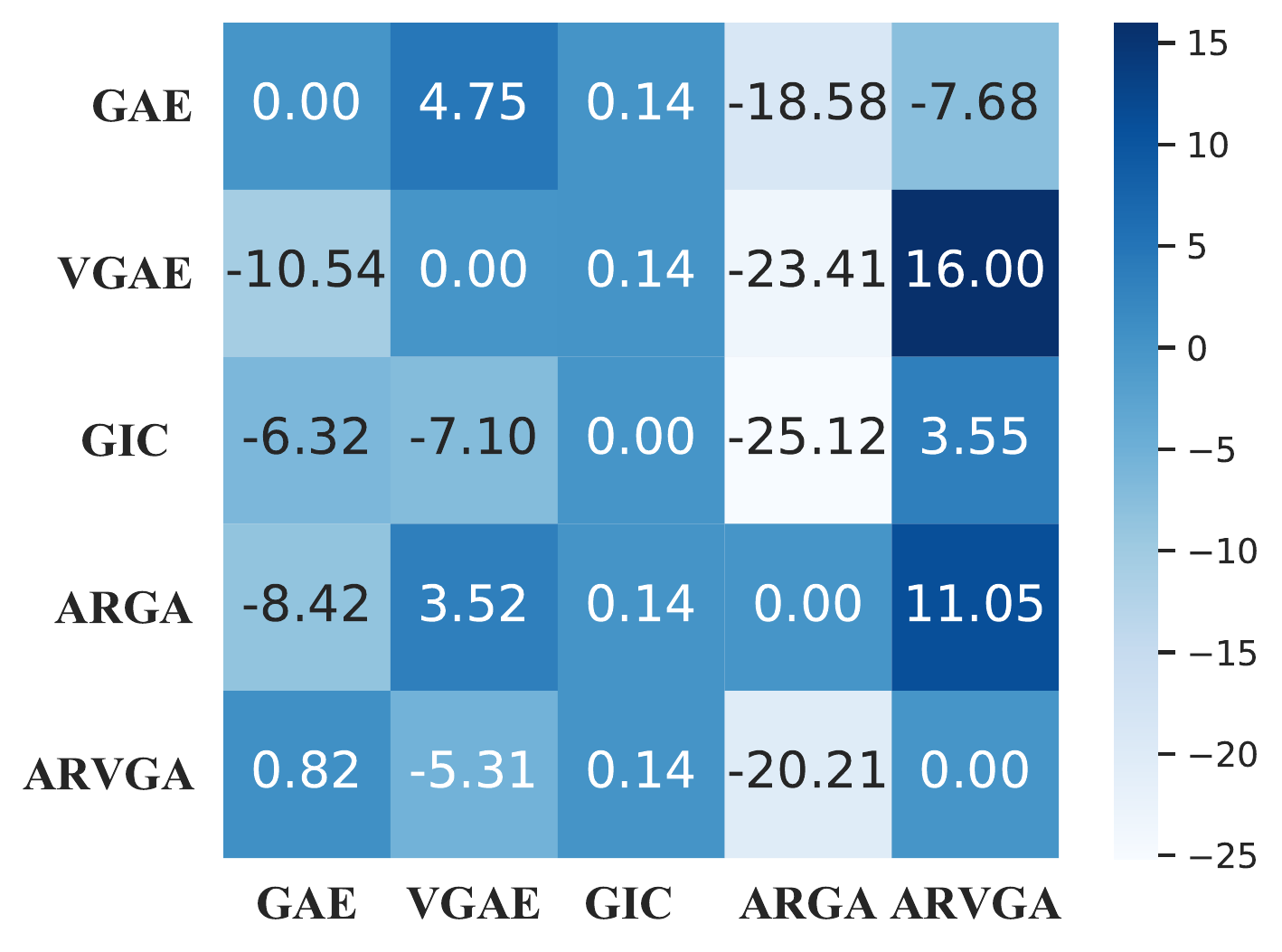}
	}
	
	\quad
	\subfigure[Pubmed]{
		\includegraphics[width=0.28\linewidth]{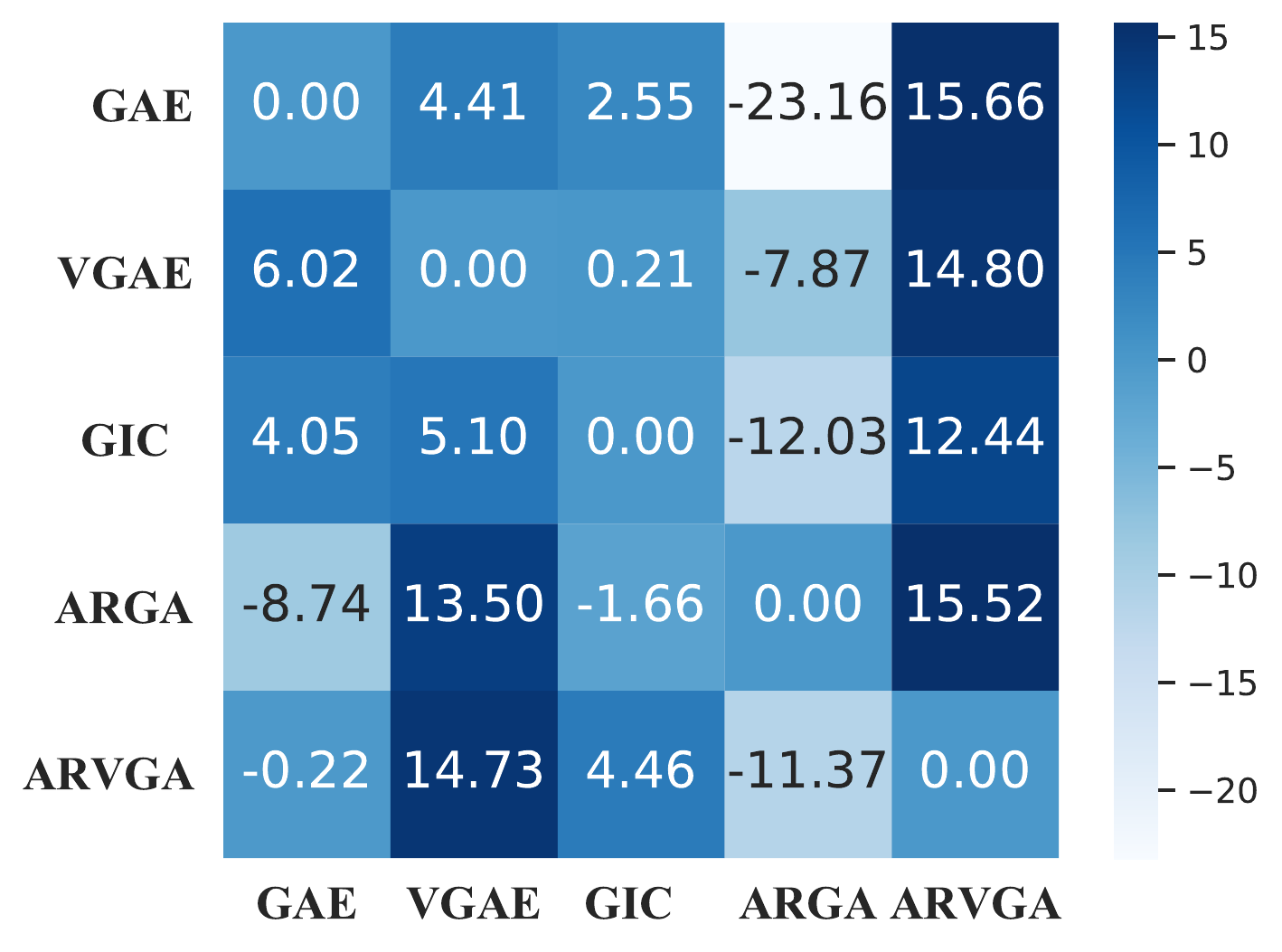}
	}
	\quad
	\subfigure[CS]{
		\includegraphics[width=0.28\linewidth]{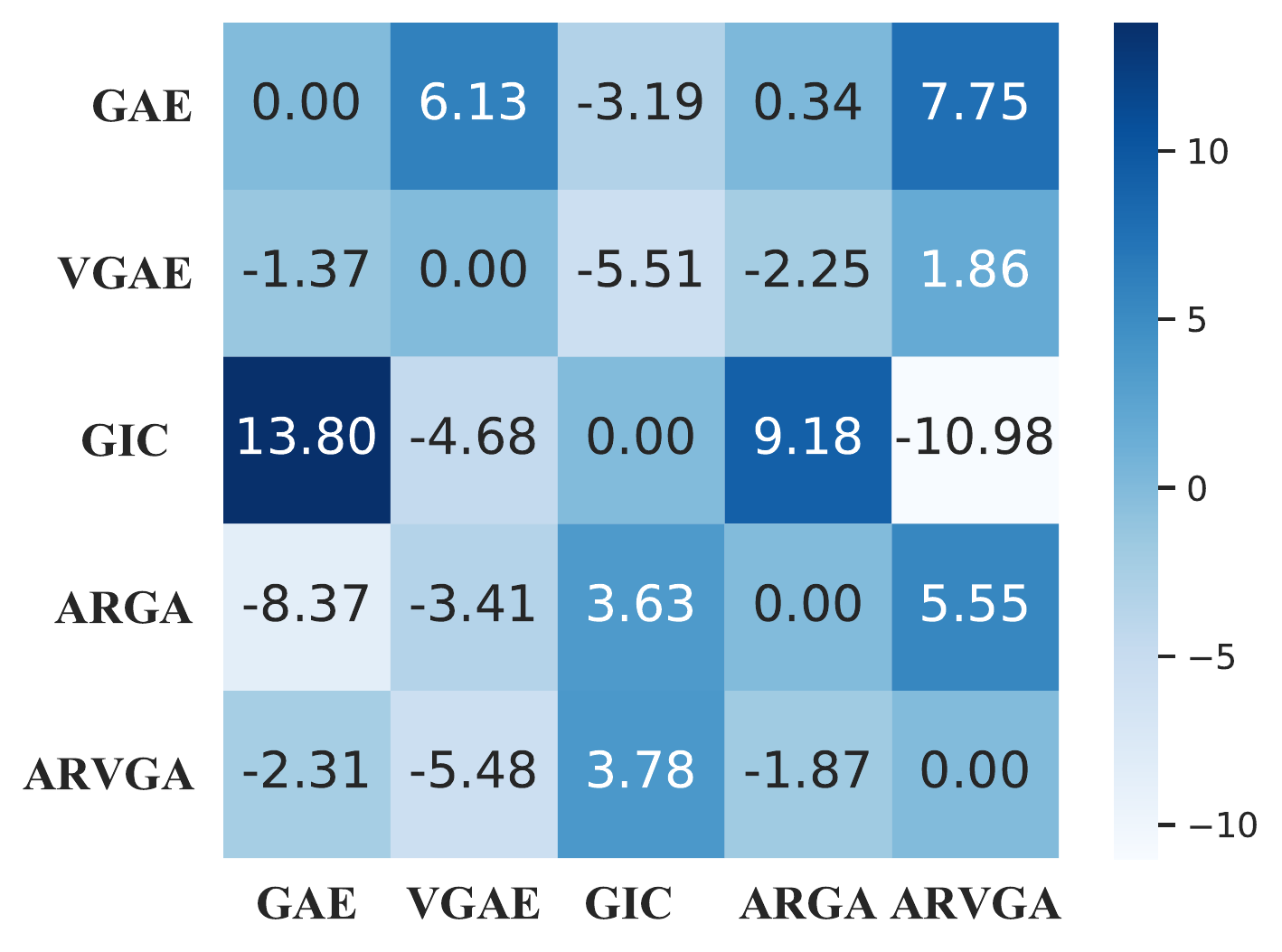}
	}
	\quad
	\caption{The ASR change heatmap under different backdoored data. We take the ASR on the diagonal as the benchmark value, 0. The ordinate is the surrogate model, and the abscissa is the target model}
	\label{fig:tra_asr}
\end{figure*}

\subsection{Performance of Link-Backdoor (RQ1)}
In the section, Link-Backdoor is conducted on five datasets and five link prediction models in white-box attack scenarios. In addition, to explore the effectiveness of nodes injection on Link-Backdoor, we conduct ablation experiments on the nodes injection module.

\subsubsection{White-box Attack of Link-Backdoor}
To evaluate the effectiveness of Link-Backdoor compared with other backdoor attack methods, we conduct attack experiments on five benchmark datasets and five SOTA link prediction methods. The attack performance is listed in Table \ref{tab:main}. To summarize, we have the following conclusions:

\noindent\textbf{Attack Performance.} Link-Backdoor achieves better backdoor attack performance. Take GIC  as an example, the ASR and AMC of the Link-Backdoor reached the highest on five datasets, and even reach $99.86\%$ on the Citeseer. Under the same settings, other methods cannot achieve this attack effect. This is because the gradient information as a guide can find more efficient structures and node features of trigger compared to other methods. Therefore, we indicate that Link-Backdoor can find more effective the trigger for attacking the target link through the strategy of gradient optimization. Besides, we found that the ASR and AMC on GIC is in most cases higher than the ASR and AMC on other models. The possible reason is that its clustering process aggregates the target links with triggers into a cluster, which enhances the attack effect of triggers. Moreover, we observed that the ASR on Citeseer is higher than the other datasets and ASR on CS is lower than other datasets. We infer that the effect of triggers is related to the edge density of the dataset, the sparser the dataset, the higher the ASR.

\noindent\textbf{Benign Performance.}  Link-Backdoor does not affect the normal performance of the target model.  in Table \ref{tab:main}, the Link-Backdoor can reach the lowest average BPD 4.14\% and can reach the minimum BPD or close to the minimum BPD in most cases. This is because the strategy of constructing triggers with injected nodes does not modify the original node features and link structure, which prevents damage to the original data. Besides, we observe that the BPD of GIC is much higher than that of other datasets. We infer that numerous injection triggers cause the GIC model to not correctly cluster nodes in the graph, resulting in a significant drop in the normal performance of the GIC model.

\subsubsection{Ablation Study of Link-Backdoor}
The strategy of nodes injection makes the Link-Backdoor more practical in real-world. In addition, we believe that the strategy of nodes injection can improve the effectiveness of backdoor attack. 

\noindent\textbf{Ablation Strategy.} To further explore the effectiveness of nodes injection, we conduct ablation experiments on the node injection module. We select nodes from the original graph to build the trigger operation to replace the node injection module, expressed as  Link-Backdoor$_{no-inj}$.

\noindent\textbf{Ablation Analysis.} As illustrated in Fig. \ref{fig:abl}, the strategy of nodes injection has a significant effect on backdoor attack. For instance, Link-Backdoor$_{no-inj}$ achieves the ASR of 68.31\% on average, i.e., in five datasets and five models, which is 14.72\% lower than that of Link-Backdoor. This can be explained as the topology of the injection nodes was only curated by Link-Backdoor, without additional redundant interference structures. It makes the target model better capture the structure and features of the triggers achieving the backdoor attack. Besides, we observe that the ASR of Link-Backdoor is significantly better than Link-Backdoor$_{no-inj}$ on GAE model. The possible reason for this is the GAE model learns the  features of triggers generated by Link-Backdoor$_{no-inj}$, but is disturbed by the features of the trigger neighbor nodes. As a result, it makes the GAE model fail to learn the features of  triggers well. This further demonstrates the effectiveness of the injection nodes strategy to launch the backdoor attack.

\begin{framed}
	For \textbf{RQ1}, from the fact that it achieved the highest average ASR (83.03\%), and the lowest average BPD (4.14\%), we can conclude that Link-Backdoor can well implement an effective and concealed backdoor attack method in link prediction by injecting nodes and gradient optimization strategies.
\end{framed}

\begin{figure}[]
	\centering
	\includegraphics[width=0.9\linewidth]{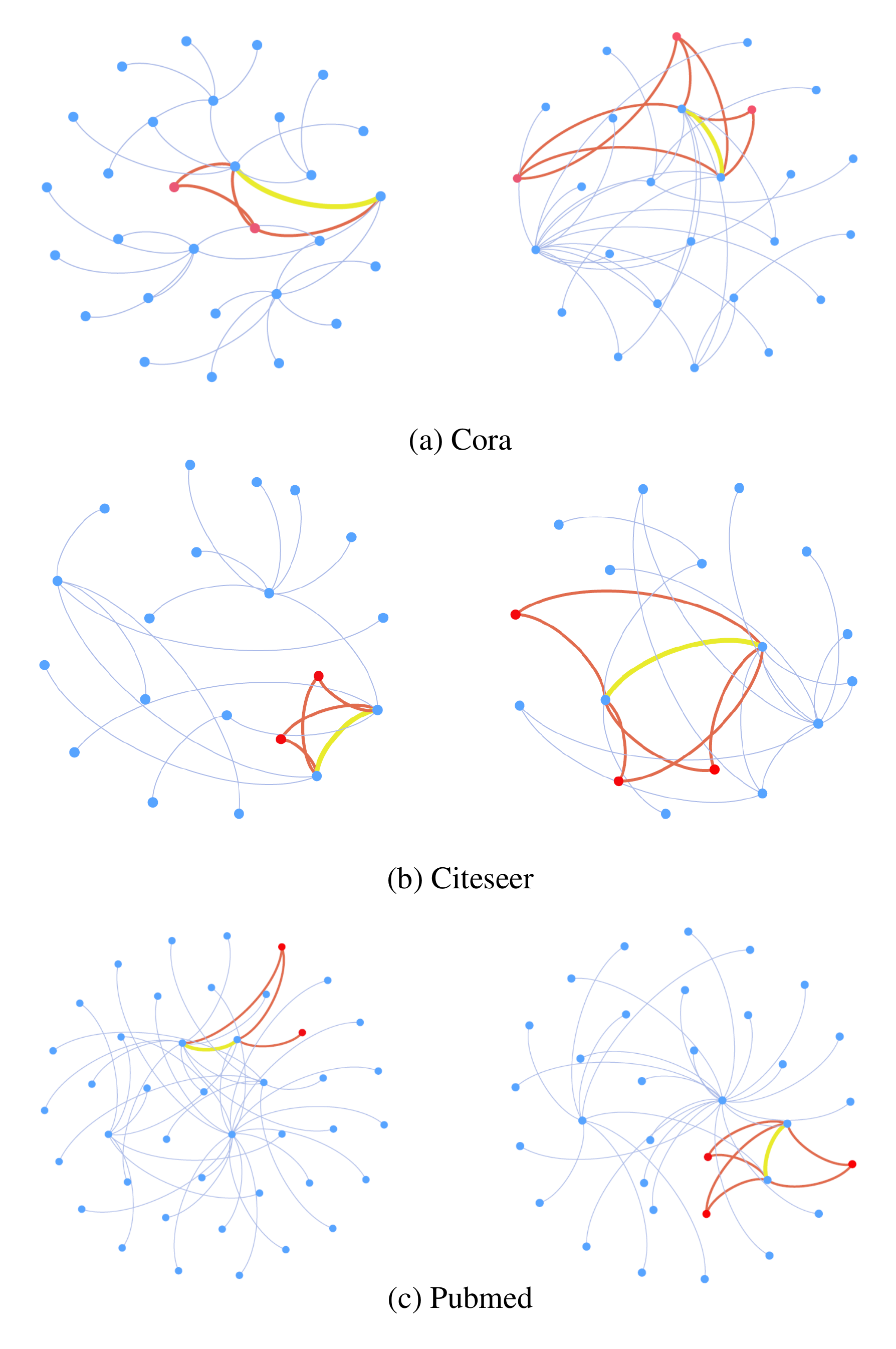}
	\caption{ The visualization results of graph with the trigger for five datasets on GAE model. The blue, red and yellow links represent the existing links in the benign graph, the links added by triggers and the target links, respectively.
	}
	\label{fig:vis}
\end{figure}

\begin{figure*}[]
	\centering
	\includegraphics[width=1\linewidth]{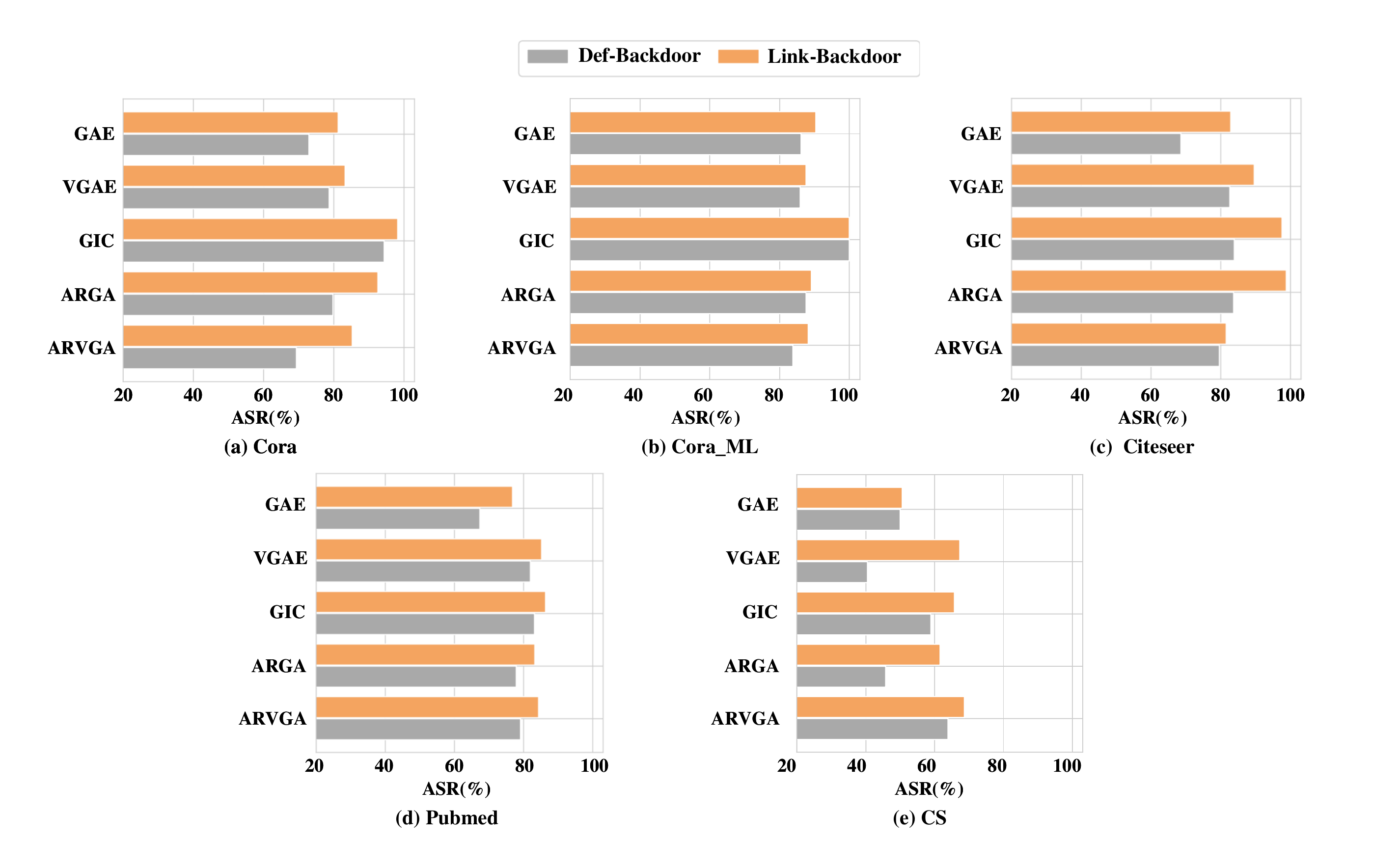}
	\caption{Defense against Link-Backdoor on five link prediction models. Link-Backdoor represents attack without defense, and Def-Backdoor represents the Link-Backdoor attack under defense.}
	\label{fig:def}
\end{figure*}

\subsection{Black-box Attack of Link-Backdoor (RQ2)}
 Due to in most practical situations, the attacker may not grasp the detail of the target link prediction model in prior, it is more practical to conduct a black-box setting attack, i.e., without any structure or parameter information of the link prediction model.
 
 \noindent\textbf{Attack Strategy.} To verify the effect of the Link-Backdoor under black-box setting, we adopt one link prediction model as the surrogate model to generate the backdoored data, and transfer the backdoored data to train other link prediction models as the target models. The results of the black-box attack experiment are shown in Fig. \ref{fig:tra_asr}.

\noindent\textbf{Attack Performance.} Link-Backdoor has a satisfactory attack effect in the black-box attack scenarios. Compared with the white-box scenario, the average ASR in the black-box scenario drops by 2.77\%. We infer that surrogate models can find efficient triggers through gradient information and injected nodes. Besides, taking VGAE as the target model of Link-Backdoor on the Cora\_ML dataset, the ASR of GAE, ARGA and ARVGA increased by 1.86\%, 0.69\%, and 3.58\%, respectively. We infer that link prediction models with good performance all have similar prediction capabilities to a certain extent. Specially, when GIC is the target model and other models are surrogate models, Link-Backdoor can still maintain the original effect or even better on GIC. We infer this phenomenon for two reasons: first, the utilization of clustering information by the GIC model improves its learning effect on structures and features of trigger, which makes GIC vulnerable to link-backdoor attacks even in black-box scenario; second, its complex clustering process results in that the gradient information cannot effectively guide the optimization of trigger, which causes its own optimized triggers that are not as effective as other surrogate models. 

\begin{framed}
	For \textbf{RQ2},  from the reduction of the average ASR is only 2.77\%, Link-Backdoor can still perform backdoor attack on the target model through the surrogate model in black-box attack scenarios, since link prediction models with good performance all have similar prediction capabilities to a certain extent.
\end{framed}

\subsection{Visualization of Link-Backdoor (RQ3)}
 We utilize gephi to visualize the graph with the trigger generated for different target links. In detail, we choose to inject two nodes and three nodes attack strategies to visualize on three datasets. The visualization results are shown in Fig. \ref{fig:vis}. The blue and red dots represent the nodes of the original network and the nodes injected by the attacker, respectively. The blue, red and yellow links represent the existing links in the clean graph, the links added by triggers and the target links, respectively.

\noindent\textbf{Visualization Analysis.} We can observe that the backdoor attack is realized through Link-Backdoor, and the cost of triggering the attack in the inference stage is relatively small, e.g., only need to inject two or three nodes and modify few links to fool the link prediction models. This shows that Link-Backdoor can find more efficient trigger structures at minimal cost by gradient information. In addition, Link-Backdoor will not destroy the related connections between nodes in the clean graph, so the graph visualization of the benign graph and the backdoored graph is similar. This means that from
an intuitive perspective, Link-Backdoor is a covert attack. 

\begin{framed}
	For \textbf{RQ3}, from the fact that the trigger only inject two or three  injection nodes, which accounts for averages 0.05\% of the entire graph on the five datasets, it can be concluded that the Link-Backdoor has a concealment.
\end{framed}



\subsection{Defense against Link-Backdoor (RQ4)}
In the section, we discuss the effectiveness of Link-Backdoor under possible defenses. Since Link-Backdoor invokes the backdoor attack through the trigger, we consider destroying the trigger information to make the backdoor attack fail in inferring stage. 

\noindent\textbf{Defense  Strategy.} To destroy the trigger information, we add noise to the features of the injection nodes in the trigger, i.e., randomly modify 10\% of the features of the injection nodes, to realize a possible defense method. The defense result on five dataset and VGAE model is shown in the Fig. \ref{fig:def}.

\noindent\textbf{Defense Performance.} Link-Backdoor is difficult to defend by modifying some features of injected nodes in the trigger. For instance, the ASR of Def-Backdoor is only 5.46\%, 7.81\% and 3.80\% lower than that of Link-backdoor on the Cora, Citeseer and Pubmed datasets respectively. This is because this defense method cannot accurately destroy the node features modified by Link-Backdoor and has no effect on the trigger's structure. This indicates the defense method, i.e., adding noise to the features of the injection nodes in the trigger, is difficult to effectively defend against Link-Backdoor.


\begin{framed}
	For \textbf{RQ4}, from the reduction of the average ASR is only 8.26\%, we can conclude that Link-Backdoor is difficult to defend by modifying some features of injected nodes in the trigger. Link-Backdoor only needs a tiny trigger to attack in the inference stage, but it is difficult to destroy this tiny trigger.
\end{framed}

\subsection{Parameter Sensitivity and Time Complexity Analysis (RQ5)}

In this section, we analyze the parameters that affect the performance of the model, and the time complexity.

\noindent\textbf{Parameter Sensitivity.} The performance of Link-Backdoor will be mainly affected by two sensitive: 1) rate of poison; 2) pre-training epoch. In the following, we will investigate their influences on the Link-Backdoor performance.

\begin{figure}[htbp]
	\centering
	\includegraphics[width=1\linewidth]{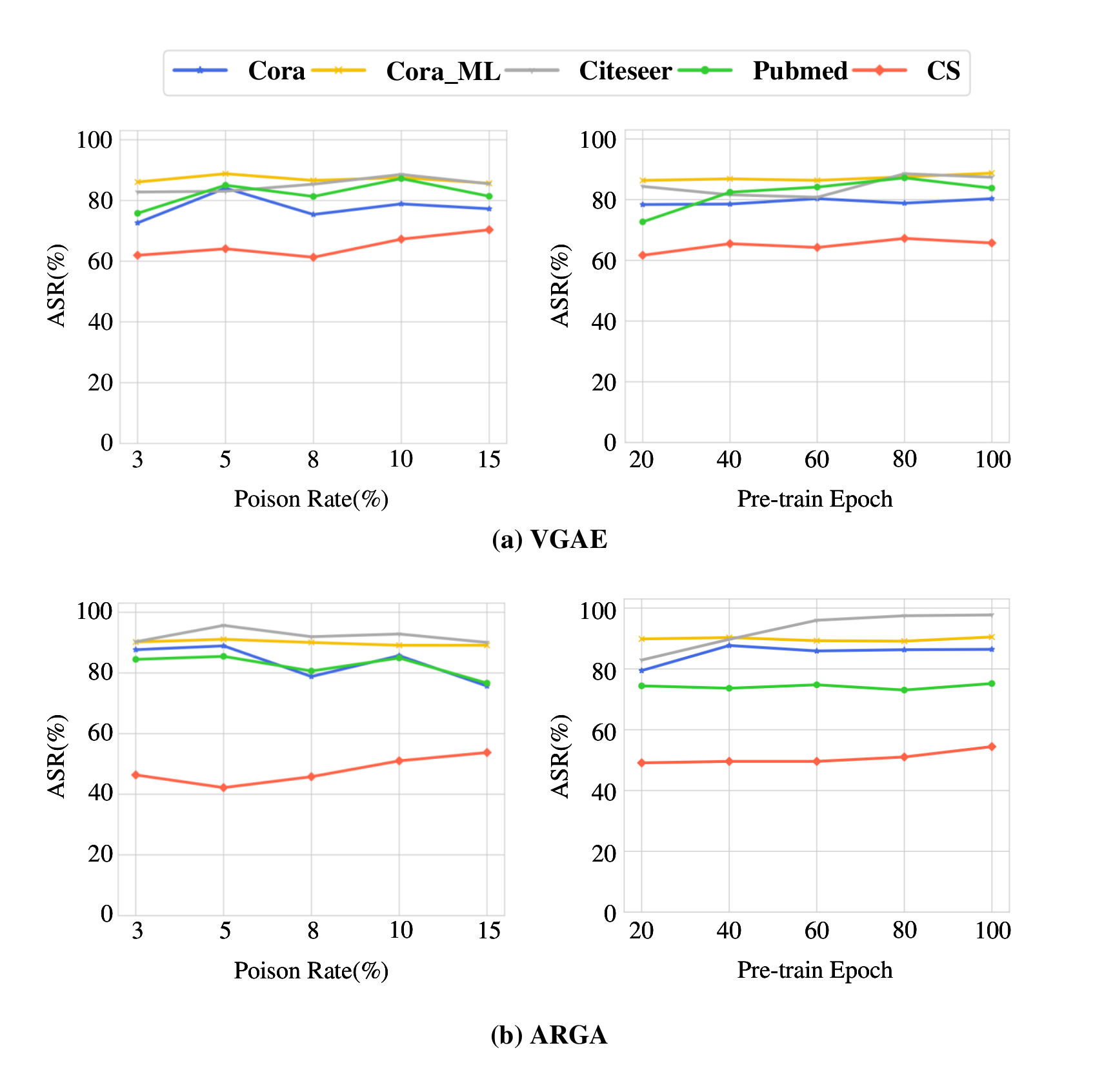}
	\caption{Hyperparameter sensitivity analysis.}
	\label{fig:param}
\end{figure}

\emph{1)Rate of Poison:} The result shown in Fig. \ref{fig:param} are obtained by the Link-Backdoor on the VGAE and ARGA model. We test the ASR with different number of target link in the training stage to measure its effect on the attack. Fig. \ref{fig:param} shows that, except for CS dataset, the attack effect of other datasets do not simply increase with the increase of the poisoning rate, but reaches a maximum value around 5\% and 10\%. This may be because the CS dataset is denser, which leads to these poisoning rates do not make the Link-Backdoor optimal, while for the other four sparser datasets, these poisoning rates are sufficient to launch an effective attack on the target model.

\emph{2)Pre-train Epoch:} Since the quality of trigger optimization is extremely dependent on whether the pre-trained model can make predictions correctly. So we tested the ASR of different pre-training time on the VGAE and ARGA model. As shown in  Fig. \ref{fig:param}, we can see that when the model training epoch is too small, the effect of the backdoor attack using the trigger generated by this pre-training model is poor, because the pre-training model training is not perfect enough, and it cannot effectively perform link prediction. The optimization of triggers does not give a valid indication. 

\begin{table}[htb]
	\caption{Time complexity to update once for different backdoor attack methods on five datasets.}
	\setlength{\tabcolsep}{0.3mm}
	\centering
	\renewcommand\tabcolsep{5.0pt}
	{
	    \resizebox{\linewidth}{!}{ \Huge
		\begin{tabular}{c|ccccc}
			\toprule \hline
			\multirow{2}{*}{Datasets}      &\multicolumn{5}{c}{Time Per Iteration (s)}\\ \cline{2-6}
			&ERB &GTA &R-Backdoor &P-Backdoor  &Link-Backdoor	
			\\ \hline
			Cora  &0.035 &30.195 &0.045  &19.064 &0.441\\
			Cora\_ML  &0.044 &64.359&0.113 &39.596 &0.959 \\
			Citeseer    &0.056 &82.457&0.131 &64.105 &0.747 \\
			Pubmed   &2.069 &5928.867 &0.189 &505.537 &16.637 \\
			CS  &1.687 &5181.851 &1.343 &864.833 &17.173\\
			
			\hline
			\bottomrule
			
		\end{tabular}}
	}
	\label{tab:time}
\end{table}

\noindent\textbf{Time Complexity Analysis.} The time cost of Link-Backdoor mainly comes from three parts, including the time cost($T_ {ptrain}$) for pre-training the model. The time cost($T_{opt}$) to optimize the features and structure of trigger, and the time cost ($T_{ptrain}$) for the final model training. Therefore, the time complexity of Link-Backdoor is:
\begin{equation}
	\begin{array}{c}
		\mathcal{O}(T_{ptrain})+\mathcal{O}(T_{opt})+\mathcal{O}(T_{train})  \sim \mathcal{O}(L),
	\end{array}
\end{equation}
where $\mathcal{O}(T_{ptrain})$ and $\mathcal{O}(T_{train})$ are depended on the types of link prediction models and the size of the training dataset. $\mathcal{O}(T_{opt})$ is the complexity of the time of trigger optimizing. $\mathcal{O}(T_{all})$ is the time complexity of Link-Backdoor.

\begin{figure}[htb]
	\centering
	\includegraphics[width=0.6\linewidth]{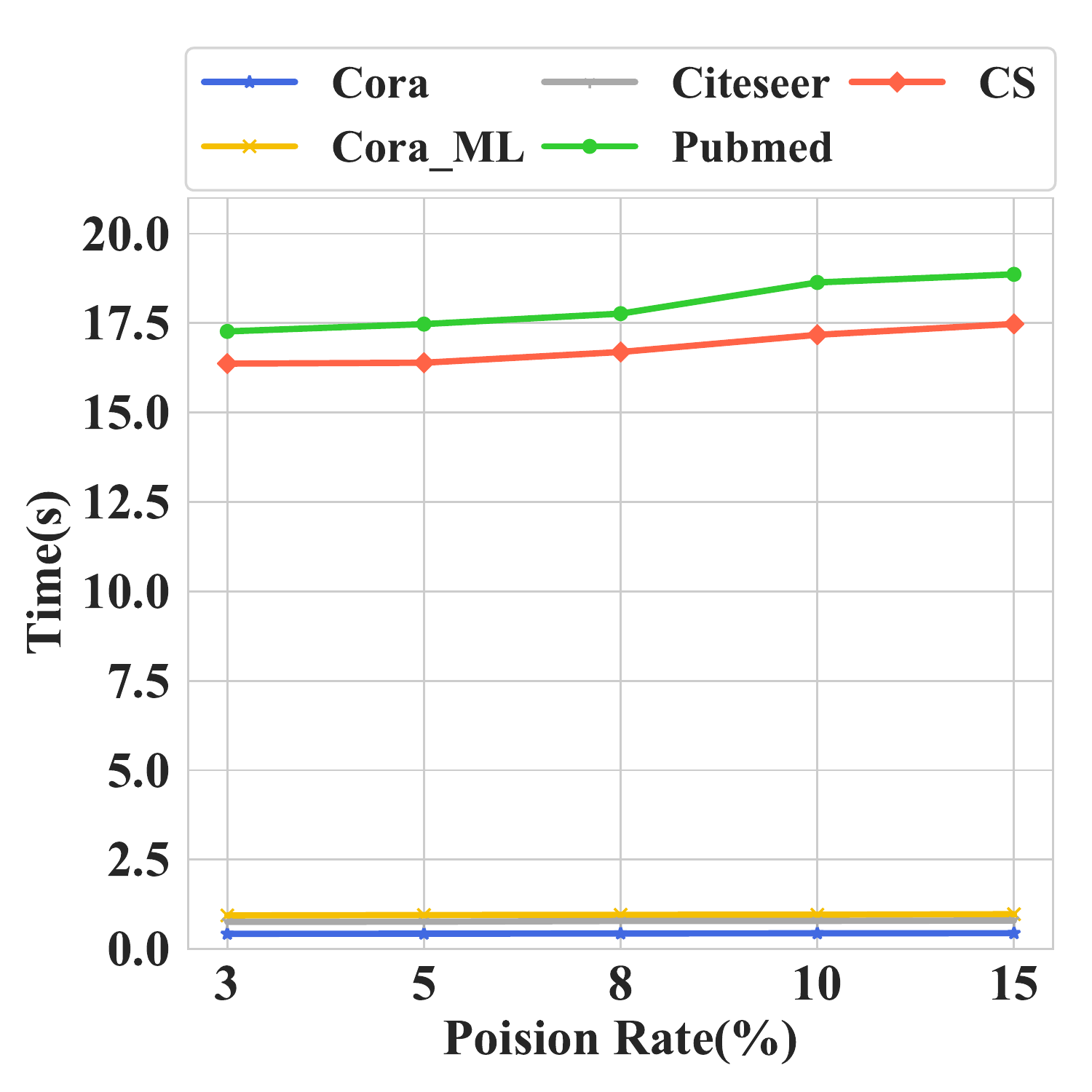}
	\caption{Time complexity for Link-Backdoor under different poisoning ratios.}
	\label{fig:time}
\end{figure}

Further, the running time of is tested. As analyzed above, since  ($T_{opt}$) and ($T_{ptrain}$) are depended on the types of link prediction models and the size of the training dataset, and not affected by backdoor attack methods. Therefore, we only measure the running time of trigger optimization for four baselines and the running time  under different poisoning ratios, and the results are shown in Fig. \ref{fig:time} and Table \ref{tab:time}. Table \ref{tab:time} shows that the time spent by P-Backdoor and GTA are much greater than that of other three methods. This is due to the fact that P-backdoor and GTA requires many iterations to find the optimal value and train the trigger generator, respectively, while other methods only need one. We observe that the time complexity of Link-Backdoor increases with the size of the dataset, while the rate of poisoning does not have much effect on the running time. This means that we can modify the poisoning ratio at will without spending too much extra time.

\begin{framed}
	For \textbf{RQ5}, Link-Backdoor requires only a few links, e.g., poisoning ratio is 3\%,with triggers to allow models to be left backdoored. Moreover, Link-Backdoor optimizes triggers by gradient, and runs much faster than P-Backdoor and GTA method.
\end{framed}


\subsection{Link-Backdoor on Non-GNNs methods (RQ6)}
Some link prediciton methods other than on GNNs-Based ones, i.e., similarity-based methods \cite{newman2001clustering}, \cite{2009Predicting}, path-based methods\cite{katz1953new}, \cite{2010Link}, embedding-based methods \cite{perozzi2014deepwalk},  \cite{grover2016node2vec}, are
also popular in practical applications. To explore the effectiveness of Link-Backdoor on non-GNNs methods, we chose two classic embedding-based methods, i.e., Deepwalk \cite{perozzi2014deepwalk} and node2vec \cite{grover2016node2vec}, to conduct backdoor attacks. The embedding-based methods are combined with MLP to complete link prediction result. The experimental results are shown in the Table \ref{tab: nonGNN}.

\begin{table}[htb]
	\caption{Link-Backdoor against non-GNNs methods.}
	\setlength{\tabcolsep}{0.3mm}
	\centering
	\renewcommand\tabcolsep{5.0pt}
	{
	    \resizebox{\linewidth}{!}{ \Huge
		\begin{tabular}{c|c|c|c|c|c}
		\toprule \hline

			 Target model &Dataset  &ASR(\%) &AMC($\times 10^{-2}$) &AUC(\%) &Benign AUC(\%)	
			\\ \hline
			  \multirow{3}{*}{Deepwalk }
			  &Cora  &52.99 &59.15 &62.60 &87.48\\
			  &Citeseer  &46.74  &59.84 &66.71 &87.93\\
			  &Pubmed &43.36 &62.53 &74.06 &86.75\\
			  
			  \hline
			  \multirow{3}{*}{Node2vec }
			  &Cora  &53.27 &64.76 &64.76 &86.68\\
			  &Citeseer  &44.89  &69.52 &74.43 &85.11\\
			  &Pubmed &45.21 &64.31 &73.95 &88.48\\

			\hline
			\bottomrule
			
		\end{tabular}}
	}
	\label{tab: nonGNN}
\end{table}

\noindent\textbf{Attack Performance.} The average ASR of the link backdoor on the deepwalk and node2vec models reaches 47.69\% and 47.79\% respectively on the Cora, Citeseer and Pubmed dataset. According to the experimental results, we can find that the link backdoor can also be attacked on the non-GNNs method, although its attack effect is not as good as that on the GNNs-Based methods. The possible reason is that non-GNNs do not utilize the feature information of nodes in the link prediction process, so that these model cannot effectively learn the node features of triggers. Besides, we find that the benign performance of non-GNNs methods degrades more on normal data than GNNs-based methods. We deduce that this is because non-GNNs methods will be more dependent on the structure of the network than GNNs-Based methods, which leads to the structure of trigger to have an impact on the original structure of the graph, and causes a benign performance degradation. Although the attack performance is not as good as GNNs-based methods on non-GNNs methods, Link-Backdoor can still achieve attacks on non-GNNs methods. 

\begin{framed}
	For \textbf{RQ6}, redfrom the fact that Link-Backdoor achieved the average ASR (47.74\%) on non-GNNs methods, we can conclude that Link-Backdoor can  implement attacks on non-GNNs link prediction methods by gradient optimization strategies.
\end{framed}

\section{Conclusions\label{Cons}}
This work represents an in-depth study on the vulnerabilities of link prediction models to backdoor attack. We propose Link-Backdoor, a novel backdoor attack framework on link prediction via nodes injection. Link-Backdoor builds a link between any two nodes through the trigger, i.e., a well-designed subgraph. Moreover, we adopt the strategy of injecting nodes to build the trigger without modifying existing data, which meets the restrictions of attack in real-world scenarios. Then, we utilize the gradient information generate by the link prediction model as a guide to optimize the structure  of the trigger and the features of the injected nodes. Extensive experiments on five benchmark datasets and five effective link prediction models demonstrate that the proposed method achieves state-of-the-art backdoor attack performance.

Link-Backdoor can currently deceive the link prediction model so that the state of the target link is predicted to exist, but it cannot make the target link unpredictable. Moreover, it is necessary to further pay attention to effective defense strategies for backdoor attacks on link prediction. The influence of trigger structure on backdoor attacks and the interpretability of backdoor attacks are also interesting research. ~\\


%
%
\section*{Compliance with Ethical Standards}

\textbf{Ethical approval:} This article does not contain any studies with human participants or animals performed by any of the authors. \\
\textbf{Conflict of Interest:} The authors  declare that they have no conflict of interest.

\bibliographystyle{spmpsci}      
\bibliography{refer}   


\section*{Appendix}
\appendix

\section{Theoretical Analysis on Link-Backdoor\label{section:theor}}
To illustrate the feasibility of backdoor attacks, we conduct a certain theoretical analysis on link-Backdoor. We unify and simplify the model structure of link prediction methods, which consists of two GCN layer and a decoding layer to facilitate formula reasoning,
\begin{equation}
    \begin{array}{c}
	  Z=f(A,X)=\Tilde{A}ReLU(\Tilde{A}XW_0)W_1 \\
	  \widehat{A}=Sigmoid(ZZ^T) 
    \end{array}\\
\end{equation}
where $\widehat{A}$ is the reconstructed graph, $W_0$ and $W_1$ are weight matrices. $\Tilde{A}=D^{-\frac{1}{2}}AD^{-\frac{1}{2}}$  is the symmetrically normalized adjacency matrix, and $D$ is the degree value matrix. 

We use the inner product function to as the encoder layer. We use $X^{\prime}$ instead of input $X$ and $A^{\prime}$ instead of input $A$ .

\begin{equation}
    \begin{array}{c}
        Z^{\prime}=f(A^{\prime},X^{\prime})=\Tilde{A}^{\prime}ReLU(\Tilde{A}^{\prime}X^{\prime}W_0)W_1 \\
    \end{array}
\end{equation}
where $A^{\prime}=A+g_A$ and $X^{\prime}=X+g_X$ , $A$ and $X$ are the adjacency matrix and feature matrix of the benign graph respectively, $g_A$ and $g_X$ are the adjacency matrix and feature matrix of the trigger respectively. The output obtained by the link prediction model can be expressed as,
\begin{equation}
	\begin{gathered}
              \widehat{A}^{\prime}=Sigmod(Z^{\prime}{Z^{\prime}}^T) 
	\end{gathered}
\end{equation}
where, $\widehat{A}^{\prime}$ represents the adjacency matrix of the graph predicted by the model. 

We uniformly use the mean square error to express the loss function of the model optimization. Then use gradient descent to update the weight parameters in the model. The update process of $W_0$ is as follows,
\begin{equation}
	\begin{gathered}
		E=\frac{1}{2}\left(\widehat{A}^{\prime}-A_T\right)^{2}\\
	\end{gathered}
\end{equation}
where $\widehat{A}^{\prime}$ represents the adjacency matrix at time $t$ of the graph predicted by the model. $A_T$ represents the adjacency matrix at time $t$ of the graph set by the attacker.  

Parameters are optimized according to the chain method,
\begin{equation}
	\frac{\partial E}{\partial W_0}=\frac{\partial E}{\partial \widehat{A}^{\prime}} \frac{\partial \widehat{A}^{\prime}}{\partial Z^{\prime}} \frac{\partial Z^{\prime}}{\partial W_0}\\
\end{equation}
where $E$ is the loss function of model training, $\widehat{A}^{\prime}$ represents the adjacency matrix of the graph predicted by the model. $W_0$ is the weight of the encoding layer. Then we derive the derivation of several parameters separately.
\begin{equation}
	\begin{gathered}
		\frac{\partial E}{\partial \widehat{A}^{\prime}}=\widehat{A}^{\prime}-A_T \\
		\frac{\partial \widehat{A}^{\prime}}{\partial Z^{\prime}}=\sigma(Z^{\prime}{Z^{\prime}}^T)(1-\sigma(Z^{\prime}{Z^{\prime}}^T))({Z^{\prime}}^T+Z^{\prime}Z_A^TZ_B^T) \\
		\frac{\partial Z^{\prime}}{\partial W_{0}}=\Tilde{A}^{\prime}(\Tilde{A}^{\prime}X^{\prime})W_1\\
	\end{gathered}
\end{equation}
where $\sigma$ is the function $Sigmoid$, $Z_A$ and $Z_B$ satisfy $Z^{\prime}=Z_A {Z^{\prime}}^T Z_B$. So the change in $W_{0}$ can be expressed as,
\begin{equation}
	\begin{gathered}
		\Delta W_{0}=-\eta \frac{\partial E}{\partial W_{0}}
		\\  =\eta (\widehat{A}^{\prime}-A_T)\sigma(Z^{\prime}{Z^{\prime}}^T)(1-\sigma(Z^{\prime}{Z^{\prime}}^T))({Z^{\prime}}^T+\\Z^{\prime}Z_A^TZ_B^T) (\Tilde{A}^{\prime}(\Tilde{A}^{\prime}X^{\prime})W_1)
	\end{gathered}
\end{equation}

The coefficient $\eta$ is the learning rate. $\Delta W_{0}$ represents the amount of change in weight $w^{0}$. The same parameters change can be obtained,
\begin{equation}
	\begin{gathered}
		\Delta W_{1}=-\eta \frac{\partial E}{\partial w^{1}}
		\\=\eta(\widehat{A}^{\prime}-A_T)\sigma(Z^{\prime}{Z^{\prime}}^T)(1-\sigma(Z^{\prime}{Z^{\prime}}^T))({Z^{\prime}}^T+\\Z^{\prime}Z_A^TZ_B^T) (\Tilde{A}^{\prime}(\Tilde{A}^{\prime}X^{\prime}W_1))\
	\end{gathered}
\end{equation}
where $\Delta W_{1}$ represent the amount of change in weight $W_{1}$. We can see that through the optimization of the trigger, the update of the model parameters is controlled. In other words, the attacker can manipulate the parameters of the model through carefully designed triggers to leave a backdoor. This also provides feasible theoretical support for backdoor attacks on link prediction methods.

\end{document}